\documentclass[pre,showpacs,showkeys,preprintnumbers,amsmath,amssymb,superscriptaddress,twocolumn]{revtex4}
\pdfoutput=0 
\usepackage{graphicx}
\usepackage{amsfonts}
\usepackage{url}
\usepackage{epstopdf}
\usepackage{color}
\usepackage{multirow}

\def\ve{\varepsilon}

\def\pa{\partial\Omega}
\def\P{{\mathbb P}}
\def\R{{\mathbb R}}

\def\U{{U}}

\def\p{s}

\begin{document}

\title{The escape problem for mortal walkers}

\author{D.~S.~Grebenkov}
 \email{denis.grebenkov@polytechnique.edu}
\affiliation{
Laboratoire de Physique de la Mati\`{e}re Condens\'{e}e (UMR 7643), \\ 
CNRS -- Ecole Polytechnique, University Paris-Saclay, 91128 Palaiseau, France}

\author{J.-F. Rupprecht}
\affiliation{
Mechanobiology Institute, National University of Singapore, \\
5A Engineering Drive 1, 117411, Singapore}

\date{\today}

\begin{abstract}
We introduce and investigate the escape problem for random walkers
that may eventually die, decay, bleach, or lose activity during their
diffusion towards an escape or reactive region on the boundary of a
confining domain.  In the case of a first-order kinetics (i.e.,
exponentially distributed lifetimes), we study the effect of the
associated death rate onto the survival probability, the exit
probability, and the mean first passage time.  We derive the upper and
lower bounds and some approximations for these quantities.  We reveal
three asymptotic regimes of small, intermediate and large death rates.
General estimates and asymptotics are compared to several explicit
solutions for simple domains, and to numerical simulations.  These
results allow one to account for stochastic photobleaching of
fluorescent tracers in bio-imaging, degradation of mRNA molecules in
genetic translation mechanisms, or high mortality rates of spermatozoa
in the fertilization process.  This is also a mathematical ground for
optimizing storage containers and materials to reduce the risk of
leakage of dangerous chemicals or nuclear wastes.
\end{abstract}





\maketitle

\section{Introduction}

The safe long-term storage of dangerous species is of paramount
importance for chemical industries, nuclear waste containers and
landfills, and military arsenals.  In spite of numerous efforts to
improve containers for dangerous chemicals or nuclear wastes, a
complete isolation is not realistic because of eventual defects and
material degradation in time.  If the species remain active forever,
their leakage is certain and is just a matter of time.  In such
situation, the quality of isolation can be characterized by the
survival probability for diffusing species to remain inside a
confining domain up to time $t$.  This is an example of the first
passage time (FPT) problems that have attracted much attention during
the last decade
\cite{Redner,Metzler,Benichou14,Holcman14}, with numerous chemical
\cite{Hanggi90}, biological
\cite{Zwanzig91,Grigoriev02,Holcman04,Bressloff13} and ecological
applications ranging from diffusion in cellular microdomains
\cite{Schuss07} to animal foraging strategies \cite{Viswanathan99}.
Most analytical results were obtained for the mean first passage time
(MFPT) through a small escape region, also known as the narrow escape
problem
\cite{Ward93,Singer06a,Singer06b,Singer06c,Pillay10,Cheviakov10,Cheviakov12,Holcman14}.

In this paper, we introduce and discuss an important extension of the
escape problem to ``mortal'' walkers.  In fact, a finite lifetime of
diffusing species is a typical situation for many biological, chemical
and ecological processes: (i) an animal or a bacterium should remain
alive while searching for food; (ii) in bio-imaging techniques,
progressive extinction of the fluorescence signal can be either due to
degradation of the tagged protein through an enzymatic reaction
(i.e. finding the target) or due to bleaching (i.e. finite
fluorescence lifetime); the latter mechanism should be taken into
account for reliable interpretation of such measurements; (iii) in
order to trigger translation, messenger RNA should not be degradated
before reaching a ribosome \cite{Eden2011}; (iv) in spite of a very
high mortality rate, the spermatozoa that search for a small egg in
the uterus or in the Fallopian tubes, need to remain alive to complete
the fertilization \cite{Alvarez14,Meerson15b,Yang16}; (v) molecules
should remain active or intact before reaching a reactive site on the
surface of a catalyst; (vi) protective materials may trap, bind or
deactivate dangerous species via a bulk reaction before they leak
through defects in the boundary of the container; (vii) for a safe
storage of nuclear wastes, the motion of radioactive nuclei should be
slowed down enough to ensure their disintegration or at least to
reduce the amount of released nuclei, etc.  While some first passage
problems have been recently extended to mortal walkers (see
\cite{Abad10,Abad12,Abad13,Yuste13,Meerson15a,Meerson15b} and references
therein), the effect of a finite lifetime of a walker onto the escape
through the boundary of two- and three-dimensional confining domains
has not been investigated.  Since the escape is not certain, because
of a possible ``death'' of the walker, the contribution of long
trajectories towards the escape region can be greatly reduced, thus
strongly affecting the conventional results.

The paper is organized as follows.  In Sec. \ref{sec:def}, we
formulate the general escape problem for mortal walkers and we then
discuss the most relevant case of a first-order bulk kinetics.  We
also introduce the exit probability of mortal walkers.  In
Sec. \ref{sec:main}, we quantify the impact of the finite lifetime of
the walkers onto their survival probability, MFPT, and exit
probability.  For this purpose, we derive the upper and lower bounds
of the Laplace-transformed survival probability and related quantities
for arbitrary bounded domains, and analyze their asymptotic behavior
at small, intermediate, and large death rates.  To check the quality
of these general estimates and asymptotics, we compare them to exact
solutions that we obtain for concentric domains and for an escape
region on the boundary of a disk.  In Sec. \ref{sec:discussion},
various extensions and applications of these results are discussed, in
particular, the problem of leakage control and optimization.  Section
\ref{sec:conclusion} summarizes and concludes the paper.

\section{Notations and equations}
\label{sec:def}

In mathematical terms, the first passage time $\tau$ is a random
variable which is determined by the survival probability $S(t;x_0) =
\P_{x_0}\{ \tau > t\}$ that a particle started at a point
$x_0\in\Omega$ has not left a confining domain $\Omega \subset \R^d$
through an escape region $\Gamma$ on the boundary $\pa$.  The survival
probability can be expressed through the diffusion propagator for the
escape problem, $G_t(x;x_0)$, i.e., the probability density for a
particle started at $x_0 \in \Omega$, to be at position $x$ after time
$t$.  For conventional immortal walkers, the propagator satisfies the
diffusion equation subject to mixed Dirichlet-Neumann boundary
condition:
\begin{equation}
\label{eq:propagator}
\begin{split}
\frac{\partial}{\partial t} G_t(x;x_0) - D \Delta G_t(x;x_0) & = 0 \quad (x\in \Omega),  \\ 
G_{t=0}(x;x_0) & = \delta(x-x_0) \quad (x\in \Omega),  \\
\frac{\partial}{\partial n} G_t(x;x_0) & = 0 \quad (x\in \pa\backslash \Gamma),  \\
G_t(x;x_0) & = 0 \quad (x\in \Gamma),  \\
\end{split}
\end{equation}
where $D$ is the diffusion coefficient, $\Delta$ the Laplace operator,
$\delta(x-x_0)$ is the Dirac distribution, and $\partial/\partial n$
the normal derivative \cite{Redner,Gardiner}.  The survival
probability is then
\begin{equation}
\label{eq:S_G}
S(t;x_0) = \int\limits_\Omega dx \, G_t(x;x_0) .
\end{equation}
The Laplace transform reduces the diffusion equation for the survival
probability (which follows from Eq. (\ref{eq:propagator})) to a
simpler Helmholtz equation,
\begin{equation}
\label{eq:tildeS}
\bigl[D \Delta - p\bigr] \tilde{S}(p;x_0) = -1,
\end{equation}
with the same mixed boundary condition, where tilde denotes the
Laplace-transformed survival probability:
\begin{equation}
\label{eq:Stilde_def}
\tilde{S}(p;x_0) = \int\limits_0^\infty dt \, e^{-tp} \, S(t,x_0) . 
\end{equation}
Once Eq. (\ref{eq:tildeS}) is solved, the inverse Laplace transform of
$\tilde{S}(p;x_0)$ yields the survival probability $S(t;x_0)$ in time
domain.  Note that the probability density of the FPT is
\begin{equation}
\label{eq:rho}
\rho(t;x_0) = - \frac{\partial}{\partial t} S(t;x_0), 
\end{equation}
while the mean FPT is simply $\tilde{S}(0;x_0)$ (in particular,
setting $p=0$ in Eq. (\ref{eq:tildeS}) yields the usual Poisson
equation for the MFPT \cite{Redner}).

\begin{figure}
\begin{center}
\includegraphics[width=70mm]{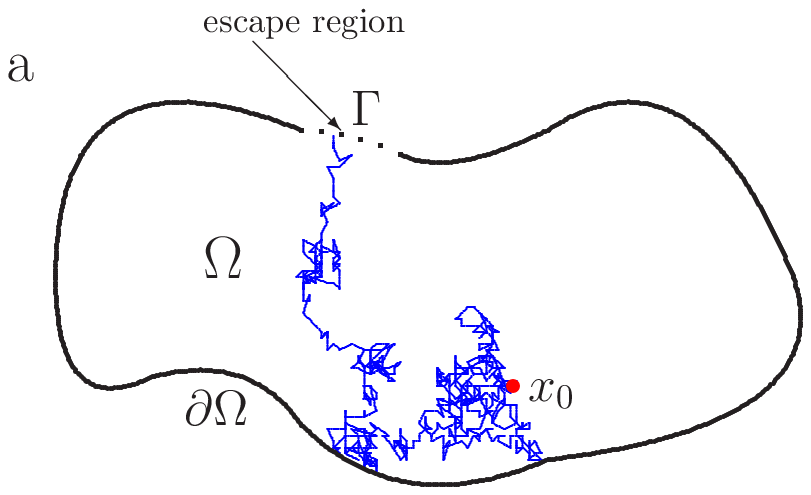} 
\includegraphics[width=80mm]{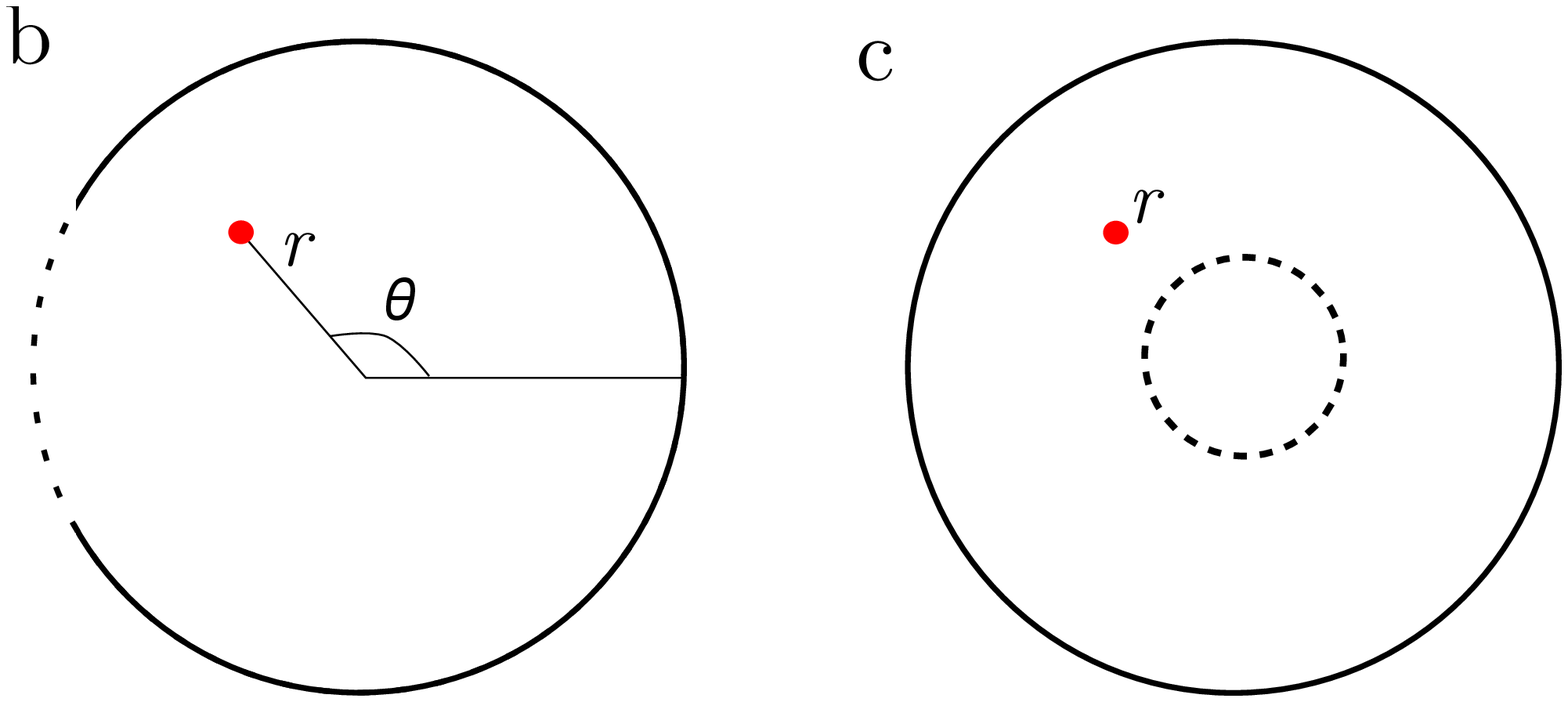} 
\end{center}
\caption{
(Color online) Schematic illustration of the escape problem.  {\bf
(a)} A general domain $\Omega\subset \R^d$ with an escape region
$\Gamma$ (dashed line) on its boundary $\pa$.  A particle is released
at a starting point $x_0$ and diffuses, with eventual reflections on
$\pa\backslash \Gamma$, until it reaches the escape region $\Gamma$.
{\bf (b,c)} Two explicitly solvable cases of the escape problem: from
a disk through an arc on its boundary {\bf (b)}, see Appendix
\ref{sec:Adisk}, and from an annulus through its inner circle {\bf
(c)}, see Appendix
\ref{sec:Aexplicit}. }
\label{fig:scheme}
\end{figure}

Next, we formulate the escape problem for mortal walkers.  We assume
that the lifetime $\chi$ of a walker is independent of the search
process.  This situation corresponds to the case of a uniform bulk
reaction or to some internal death mechanism of a walker, e.g., a life
cycle of a bacterium or a radioactive decay.  In this case, the
survival probability of a mortal walker inside the domain $\Omega$ up
to time $t$, $S_\mu(t;x_0)$, is simply the product of $S(t;x_0)$ and
$Q(t) = \P\{\chi > t\}$:
\begin{equation}
\label{eq:Smu0}
S_\mu(t;x_0) = \P\{ \min\{\tau, \chi\} > t\} = S(t;x_0)~ Q(t).
\end{equation}
While this expression is general, it remains formal as the survival
probability $S(t;x_0)$ is not known analytically except for some
elementary cases (see Appendix \ref{sec:Aexplicit}).

A particular simplification occurs in the relevant case of an
exponentially distributed lifetime, $Q(t) = \exp(-\mu t)$, with $\mu$
being the death rate (or $1/\mu$ being the mean lifetime), for which
the Laplace transform of Eq. (\ref{eq:Smu0}) reads
\begin{equation}
\label{eq:Smu}
\tilde{S}_\mu(p;x_0) = \tilde{S}(p+\mu;x_0).
\end{equation}
This is equivalent to adding the term $-\mu \tilde{S}$ into
Eq. (\ref{eq:tildeS}) to describe a first-order bulk kinetics or
radioactive decay.  The MFPT for mortal walkers is then
\begin{equation}
\label{eq:MFET}
\langle \tau_\mu \rangle = \tilde{S}_\mu(0;x_0) = \tilde{S}(\mu;x_0) ,
\end{equation} 
where $\langle \cdots\rangle$ denotes the expectation.  Since the
Laplace transform of a positive function monotonously decreases, the
MFPT monotonously decreases with $\mu$.  In particular, the MFPT for
mortal walkers is smaller than that for immortal ones.  This is
expected because long trajectories to the escape region are
progressively eliminated as $\mu$ increases.  

Finally, we introduce the exit probability of mortal walkers,
$H_\mu(x_0)$, i.e., the probability that a walker started from $x_0$
leaves the confining domain before dying: $H_\mu(x_0) = \P\{ \tau <
\chi\}$.  Since $\tau$ and $\chi$ are independent, the exit
probability can in general be expressed as
\begin{equation}
\label{eq:Hexit_def}
\begin{split}
H_\mu(x_0) & = \int\limits_0^\infty \P\{\chi > t\} \, \P\{ \tau \in (t,t+dt)\} \\
&  = \int\limits_0^\infty dt\, Q(t) \, \rho(t; x_0) . \\
\end{split}
\end{equation}
For an exponentially distributed lifetime, the exit probability becomes
\begin{equation}
\label{eq:Hexit}
H_\mu(x_0) = \int\limits_0^\infty dt\, e^{-\mu t} \, \rho(t;x_0) = 1 - \mu \tilde{S}(\mu;x_0) = 1 - \mu \langle \tau_\mu \rangle ,   
\end{equation}
where we used Eq. (\ref{eq:rho}).  The exit probability satisfies
\begin{equation}
\label{eq:Hmu_def}
\begin{split}
D\Delta H_\mu(x_0) - \mu H_\mu(x_0) & = 0  \quad (x_0\in\Omega), \\
H_\mu(x_0) & = 1 \quad (x_0\in \Gamma), \\
\frac{\partial}{\partial n} H_\mu(x_0) & = 0 \quad (x_0\in \pa\backslash \Gamma) . \\
\end{split}
\end{equation}
As expected, this probability is equal to $1$ for immortal walkers
(i.e., for the conventional escape problem with $\mu = 0$) and
monotonously decreases to $0$ as $\mu$ increases.  For mortal walkers,
the exit probability characterizes the storage safety of a container.

In many applications, the starting point $x_0$ is not fixed but
randomly distributed over the confining domain.  In this case, one
considers the global, or volume-averaged MFPT:
\begin{equation}
\overline{\langle \tau_\mu\rangle } = \frac{1}{|\Omega|} \int\limits_{\Omega} dx_0 \, \langle \tau_\mu \rangle  \,,
\end{equation}
where $|\Omega|$ is the volume of $\Omega$.  In analogy, we define the
global exit probability (GEP) to characterize the overall safety of
the container:
\begin{equation}
\overline{H}_\mu = \frac{1}{|\Omega|} \int\limits_{\Omega} dx_0 \, H_\mu(x_0).
\end{equation}
Integrating Eq. (\ref{eq:Hmu_def}) over the confining domain and using
the Green formula and boundary conditions, one gets another
representation of the GEP:
\begin{equation}
\label{eq:Hmu_other}
\overline{H}_\mu = \frac{D}{\mu |\Omega|} \int\limits_\Gamma dx_0 \, \frac{\partial H_\mu(x_0)}{\partial n} 
\end{equation}
that will used to derive its general asymptotic expression in the
limit of large death rates.

When the mean lifetime $1/\mu$ is much larger that the time needed to
find the escape region by immortal walkers, i.e., $\mu \langle \tau_0
\rangle \ll 1$, the Taylor expansion of Eq. (\ref{eq:Stilde_def}) yields
\begin{equation}
\label{eq:MFPT_small}
\langle \tau_\mu \rangle = \langle \tau_0 \rangle - \frac12 \mu \langle \tau_0^2 \rangle + O(\mu^2)
\end{equation}
and, from Eq. (\ref{eq:Hexit}),
\begin{equation}
\label{eq:Hmu_small}
H_\mu(x_0) = 1 - \mu \langle \tau_0 \rangle + \frac12 \mu^2 \langle \tau_0^2 \rangle + O(\mu^3) .
\end{equation}
The problem is reduced to the analysis of the moments of the FPT of
immortal walkers  
\begin{equation}
\label{eq:tau_moments}
\langle \tau_0^n \rangle = (-1)^n \left(\frac{\partial^n}{\partial \mu^n} H_\mu(x_0) \right)_{\mu = 0} .
\end{equation}
In particular, the exact formulas for the MFPT are known for several
simple domains
\cite{Singer06b,Caginalp12,Holcman14,Rupprecht15,Marshall16} (see also
Appendices \ref{sec:Aexplicit} and \ref{sec:Adisk}).  When the escape
region is small, the following asymptotic behavior was established
(see the review \cite{Holcman14} and references therein):
\begin{equation}
\label{eq:MFPT_classic}
\langle \tau_0 \rangle \simeq \begin{cases} \frac{|\Omega|}{\pi D} \bigl(\ln (1/\epsilon) + O(1)\bigr) \quad (d=2),\cr
\frac{|\Omega|}{4DR\epsilon} + O(\ln \epsilon) \hskip 11mm (d=3),\end{cases}
\end{equation}
where $R = |\pa|^{1/(d-1)}$ is the characteristic size of the domain,
and $\epsilon = (|\Gamma|/|\pa|)^{1/(d-1)}$ is the perimeter of the
escape region $\Gamma$ normalized by the perimeter of the boundary
$\pa$ (for $d=2$), or the square root of the area of the escape region
normalized by the area of the boundary (for $d=3$).

In the context of chemical reactions, the escape region can be
interpreted as an active catalytic site on an inert confining boundary
of a reactor, the FPT $\tau_\mu$ is the (random) reaction time (for a
perfectly reactive catalyst), $\langle \tau_\mu \rangle$ is the mean
reaction time, and $H_\mu(x_0)$ is the target encounter probability.
While we adopt the language of first-passage processes, the following
results can be easily translated into the chemical or biochemical
context.

In what follows, we investigate the Laplace-transformed survival
probability, the MFPT, the exit probability and the GEP for mortal
walkers in the limit of short lifetimes, $\mu \langle \tau_0 \rangle
\gg 1$, which turns out to be more relevant for many applications,
e.g., for developing safe containers.  Note also that the large $\mu$
limit captures the short-time asymptotic behavior of the survival
probability and related quantities according to Tauberian theorems.
In contrast to the long-time asymptotics, the short-time behavior of
first passage distributions is generally not universal \cite{Godec16}.
Nevertheless, we will obtain a universal relation (\ref{eq:Hmu_limit})
for the global exit probability within the short-time scale regime.

\section{Main results}
\label{sec:main}

\subsection{Upper and lower bounds}

First, we establish the upper and lower bounds for the
Laplace-transformed survival probability.  An elementary upper bound
follows from the trivial inequality $S(t,x_0) \leq 1$ by applying the
Laplace transform:
\begin{equation}
\label{eq:Supper}
\tilde{S}(\mu;x_0) \leq \frac{1}{\mu} .
\end{equation}
The lower bound can be obtained from the continuity of Brownian motion
that implies that the FPT to a subset $\Gamma$ of the boundary $\pa$
is greater than the FPT from the center of the ball of radius $|x_0 -
\pa|$ to its boundary, where $|x_0 - \pa|$ is the distance from $x_0$
to the boundary $\pa$.  In probabilistic terms, it means that
\begin{equation}
S(t;x_0) \geq S_{B_d(|x_0-\pa|)}(t;0), 
\end{equation}
where $S_{B_d(R)}(t;0)$ is the survival probability inside a
$d$-dimensional ball of radius $R$, $B_d(R)$, for a particle started
from the center of that ball.  We claim a stronger inequality
\begin{equation}
\label{eq:Sineq}
S(t;x_0) \geq S_{B_d(|x_0-\Gamma|)}(t; 0) ,
\end{equation}
where the distance $|x_0-\pa|$ to the boundary $\pa$ is replaced by
the distance $|x_0-\Gamma|$ to the escape region $\Gamma$ (see
Appendix \ref{sec:conjecture}).  Since the Laplace transform of
positive functions does not affect the inequality, one also gets the
explicit lower bound in the Laplace domain
\begin{equation}
\label{eq:Stilde_lower}
\tilde{S}(\mu;x_0) \geq \tilde{S}_{B_d(|x_0-\Gamma|)}(\mu; 0) .
\end{equation}
The right-hand side is known explicitly:
\begin{equation}
\label{eq:tildeS_ball0}
\tilde{S}_{B_d(R)}(\mu;0) = \frac{1}{\mu}\bigl(1 - \U_d(R\sqrt{\mu/D}) \bigr),
\end{equation}
with
\begin{equation}
\label{eq:H_upperdef}
\U_d(x) = \frac{\bigl(x/2\bigr)^{\frac{d}{2}-1}}{\Gamma\bigl(\frac{d}{2}\bigr)~ I_{\frac{d}{2}-1}(x)} ,
\end{equation}
where $\Gamma(d/2)$ is the Gamma function, and $I_\nu(z)$ is the
modified Bessel function of the first kind (see Appendix
\ref{sec:Aball}).  When $x \gg 1$, one has
\begin{equation}
\label{eq:tildeS_ball0p}
\U_d(x) \simeq \frac{\sqrt{\pi}\, 2^{\frac{3-d}{2}}}{\Gamma(d/2)} \, x^{\frac{d-1}{2}}  e^{-x} .
\end{equation}
Combining the upper and lower bounds (\ref{eq:Supper},
\ref{eq:Stilde_lower}), one concludes that the MFPT,
$\langle\tau_\mu\rangle = \tilde{S}(\mu;x_0)$, can be accurately
approximated as $1/\mu$ at large $\mu$ for any domain and any escape
region, the error of this approximation vanishing as a
stretched-exponential function according to
Eq. (\ref{eq:tildeS_ball0p}).  As a consequence, the MFPT becomes
fully controlled by the lifetime of the walker in this limit, as
expected.  Similarly, the global MFPT, $\overline{\langle
\tau_\mu\rangle}$, behaves as $1/\mu$ at large $\mu$, although it
includes contribution from points which are close to the escape
region.

\begin{figure}
\begin{center}
\includegraphics[width=80mm]{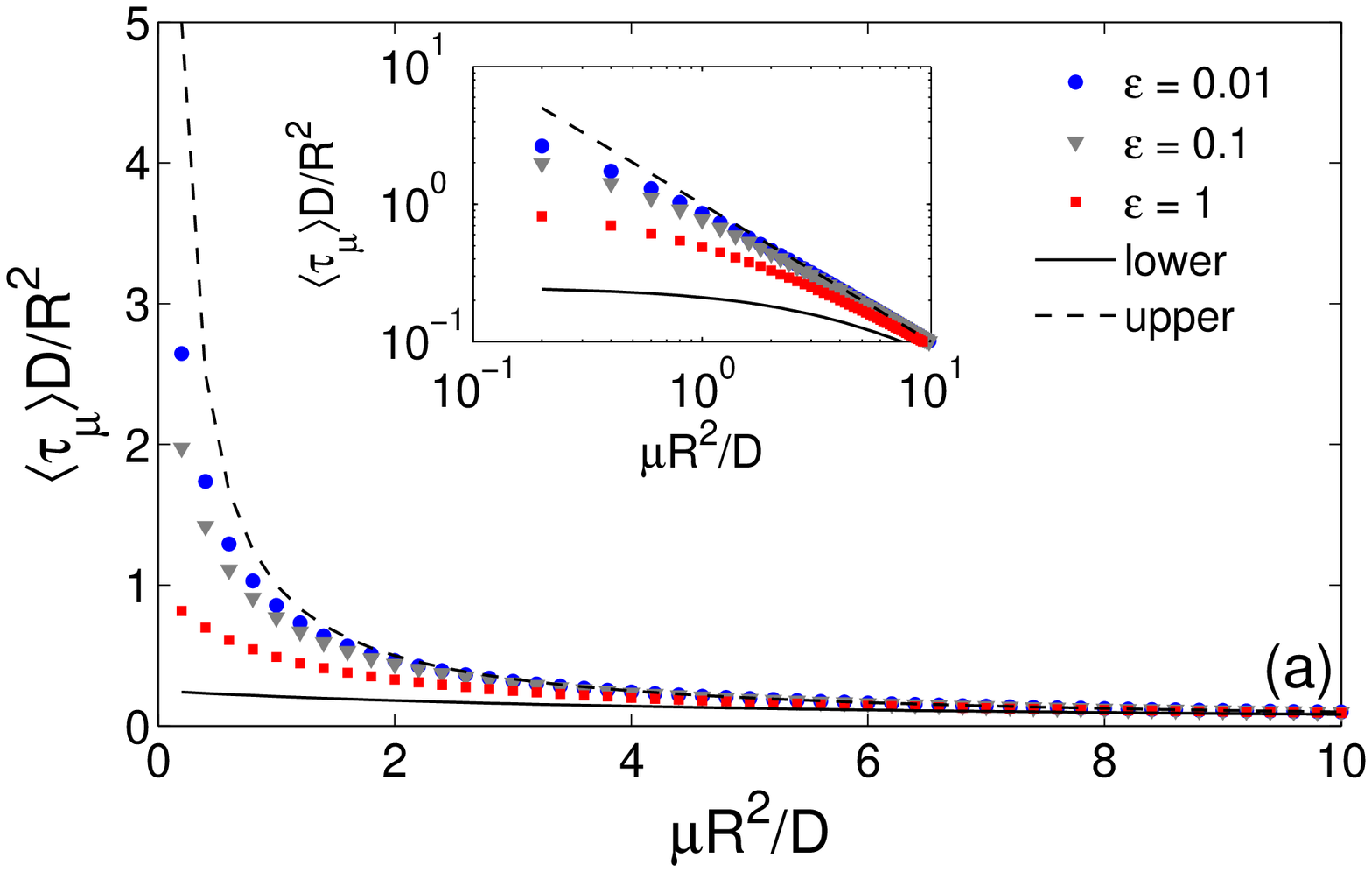} 
\includegraphics[width=80mm]{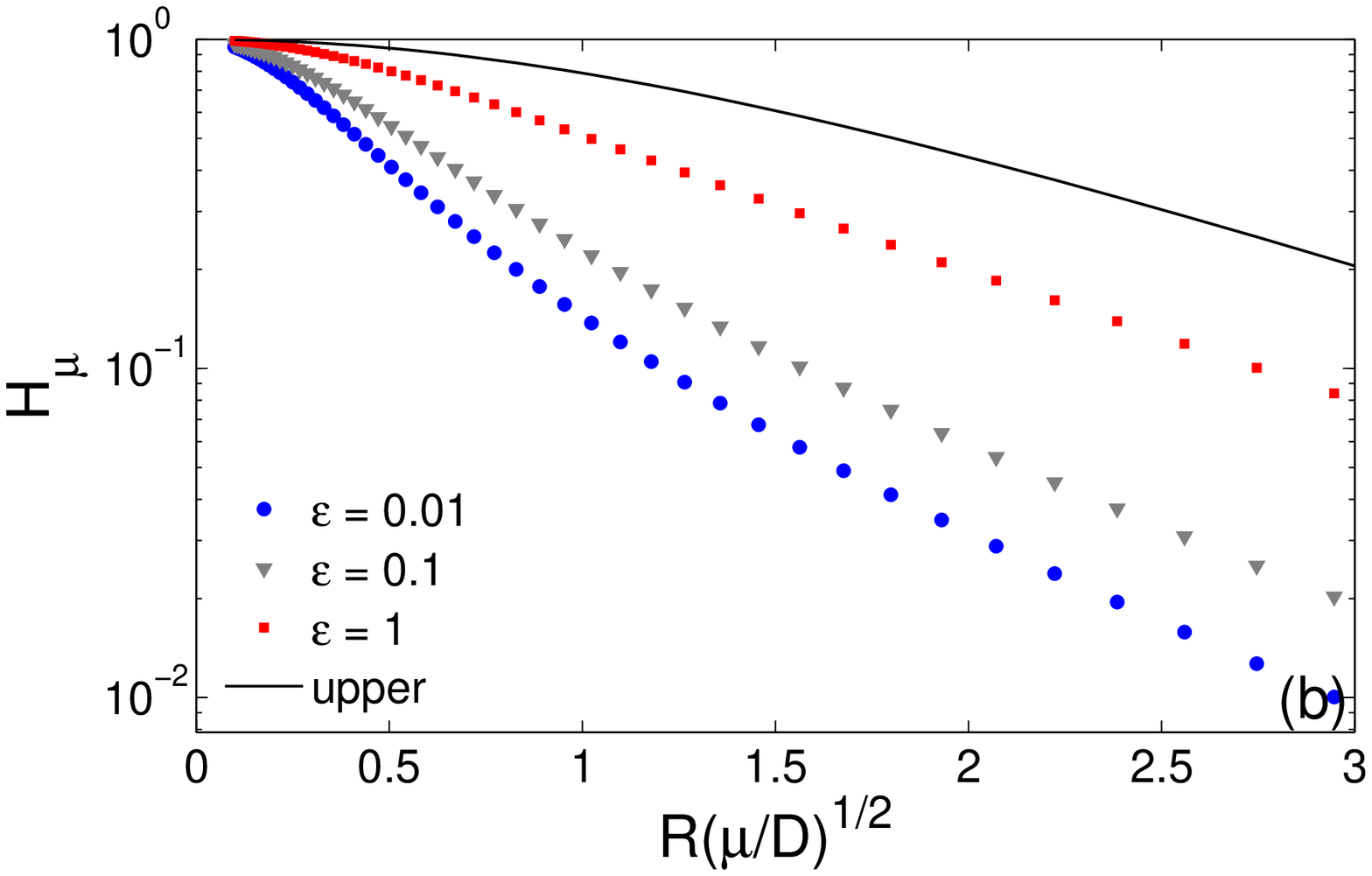} 
\end{center}
\caption{
(Color online) The MFPT $\langle\tau_\mu\rangle$ {\bf (a)} and the
exit probability $H_\mu(x_0)$ {\bf (b)}, as functions of $\mu R^2/D$
and $R\sqrt{\mu/D}$, respectively, for the disk of radius $R$ with the
escape arc $(\pi-\ve,\pi+\ve)$ on the boundary, for several $\ve$
(here $x_0 = 0$).  Inset shows the MFPT on log-log scale. For the
MFPT, the exact solution (\ref{eq:Sdisk_exact}) is compared to the
upper and lower bounds (\ref{eq:Supper}, \ref{eq:Stilde_lower}).  For
the exit probability, the exact solution (\ref{eq:Hmu_disk}) is
compared to the upper bound (\ref{eq:H_upper}) (the lower bound being
zero).  }
\label{fig:Smu_0}
%
\end{figure}

Figure \ref{fig:Smu_0}a illustrates this behavior in the case of a
disk with an escape region on its boundary, for which the exact
formula for the Laplace-transformed survival probability was recently
derived \cite{Rupprecht15} (see Appendix \ref{sec:Adisk}).  One can
see that the two bounds accurately approximate the MFPT at large
$\mu$.  In turn, they do not control the behavior of the MFPT at small
$\mu$, in which case the lower bound tends to $|x_0-\Gamma|^2/(2dD)$,
and the upper bound diverges.  In this limit, one can use the
asymptotic relation (\ref{eq:MFPT_small}).

In turn, the exit probability $H_\mu(x_0)$ becomes a nontrivial
characteristics of the escape problem at large $\mu$.  According to
Eqs. (\ref{eq:Hexit}, \ref{eq:Stilde_lower}), $H_\mu(x_0)$ is bounded
as
\begin{equation}
\label{eq:H_upper}
H_\mu(x_0) \leq \U_d\bigl(|x_0 - \Gamma|\sqrt{\mu/D}\bigr) .
\end{equation}
Since the upper bound is independent of the size of the escape region,
it does not capture well the accessibility of this region.  In fact,
the exit probability is expected to decay faster with $\mu$ for
smaller escape regions.  This is illustrated in Fig. \ref{fig:Smu_0}b
that shows $H_\mu(x_0)$ for the disk with escape regions of various
sizes.  While the upper bound over-estimates the exit probability, it
captures correctly its asymptotic decay as a stretched-exponential
function.  The quality of the upper bound (\ref{eq:H_upper}) is
further discussed in the case of a disk in Appendix \ref{sec:AHupper}.
Note also that this upper bound is equal to the exit probability in
the case of a ball with the escape region on the whole boundary and
the starting point at the origin.

According to the inequality (\ref{eq:H_upper}) and numerical analysis,
the exit probability $H_\mu(x_0)$ at large $\mu$ exponentially decays
with the distance from the starting point to the escape region.  For a
better control of this decay, one needs a lower bound for the exit
probability.  In Appendix \ref{sec:A_upper}, we partly solve this
problem for the specific case when the escape region is the whole
boundary: $\Gamma = \pa$.  Finding an appropriate lower bound for the
general escape problem (with $\Gamma \ne \pa$) remains an open
problem.

\subsection{Global exit probability}

When the starting point is uniformly distributed, some walkers start
near the escape region, and the stretched-exponential decay with $\mu$
is expected to be replaced by a slower power law.  Determining an
exact expression for the GEP appears to be a challenging problem
because of mixed boundary conditions.  Here, we derive an
approximation for the GEP for intermediate death rates, which also
turns out to be exact in the limit of small death rates.  For small
escape regions, this approximation shows satisfactory agreement with
the results of numerical simulations over a broad range of $\mu$.

The survival probability $S(t;x_0)$ in a bounded domain admits a
spectral decomposition on appropriate Laplacian eigenfunctions and
thus decreases exponentially at long times as $S(t;x_0) \sim A(x_0)
e^{-\lambda t}$, where $\lambda$ is the smallest Laplacian eigenvalue,
and $A(x_0)$ is a coefficient.  When the escape region is small as
compared to the size of the confining domain, one expects $A(x_0)
\simeq 1$ and $\overline{\langle \tau_0\rangle} = 1/\lambda$,
independently of the starting point $x_0$ if $x_0$ is far from the
escape region \cite{Ward93,Cheviakov11,Isaacson13}.  As a consequence,
the volume-averaged survival probability behaves as $\overline{S}(t)
\sim e^{-\lambda t}$ in this narrow escape limit, from which the
Laplace transform yields
\begin{equation}
\label{eq:Hmu_approx}
\overline{H}_\mu = 1 - \mu \, \overline{\tilde{S}}(\mu) \approx \frac{1}{1 + \mu \, \overline{\langle \tau_0\rangle}}.
\end{equation}
In the limit $\mu \to 0$, one retrieves $\overline{H}_0 = 1$ for any
escape region size, independently of its smallness, as expected.  In
turn, the exit probability vanishes for any $\mu > 0$ in the limit of
shrinking escape region because $\overline{\langle \tau_0\rangle}\to
\infty$.  Note also that this expression can be re-written as
$1/\overline{\langle \tau_\mu\rangle} = 1/\overline{\langle
\tau_0\rangle} + \mu$, i.e., the volume-averaged MFPT for mortal
walkers is the harmonic mean of $\overline{\langle \tau_0\rangle}$ and
$1/\mu$.  We point out that Eq. (\ref{eq:Hmu_approx}) is exact, to the
first order, in the limit $\mu \overline{\langle\tau_0\rangle} \ll 1$.
This can be seen by comparing the asymptotic expansion of
Eq. (\ref{eq:Hmu_approx}) to the integral over $x_0$ of the Taylor
expansion (\ref{eq:Hmu_small}).

\begin{figure}
\begin{center}
\includegraphics[width=80mm]{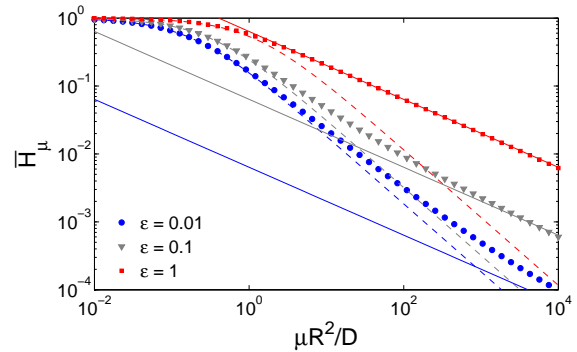} 
\end{center}
\caption{
(Color online) The global (or volume-averaged) exit probability
$\overline{H}_\mu$ (symbols) from Eq. (\ref{eq:GEP_disk}) as a
function of $\mu R^2/D$ for the disk of radius $R$ with the escape arc
$(\pi-\ve,\pi+\ve)$ on the boundary.  Dashed lines show the
approximation (\ref{eq:Hmu_approx}), with the volume-averaged MFPT
$\overline{\langle \tau_0\rangle} = \frac{R^2}{D}
\bigl(1/8 - \ln(\sin(\ve/2))\bigr)$ \cite{Singer06b} while solid lines
present the asymptotic relation (\ref{eq:Hmu_limit}).}
\label{fig:Hmean}
\end{figure}

We compare the approximation (\ref{eq:Hmu_approx}) to two exact
solutions: the concentric escape region (Appendix \ref{sec:Aexplicit})
and the disk with an escape arc $(\pi-\ve,\pi+\ve)$ on the boundary
(Appendix \ref{sec:Adisk}).  We present the results only for the
latter case, for which Fig. \ref{fig:Hmean} shows the GEP as a
function of the death rate.  The approximation (\ref{eq:Hmu_approx})
(shown by dashed lines) accurately captures the behavior of the GEP at
small $\mu$ and under-estimates its values at larger $\mu$.  In the
narrow escape configuration, when $\overline{\langle \tau_0\rangle}
\gg R^2/D$, the approximation (\ref{eq:Hmu_approx}) remains accurate
even for $\mu \sim 1/\overline{\langle \tau_0\rangle}$.  However,
Eq. (\ref{eq:Hmu_approx}) fails at large death rates $\mu \gg D/(\ve
R)^2$ that correspond to the short-time behavior of the FPT
distribution.

The mono-exponential approximation of the survival probability fails
at very short times.  Similarly to \cite{Godec16}, we show that in the
large death rate limit (which corresponds to the few-encounter regime
in \cite{Godec16}), the notion of a kinetic rate in the traditional
bulk sense does not hold, and the asymptotic regime of the GEP should
be different.  In this limit, the exit probability $H_\mu(x_0)$
rapidly decays from the escape region $\Gamma$ towards the bulk.  In
the vicinity of each escape point, the exit probability along normal
vector can be approximated by the one-dimensional solution of
Eq. (\ref{eq:Hmu_def}) on the positive half-line: $H_\mu(x_0) =
\exp(-x_0 \sqrt{\mu/D})$.  Substituting this approximation in
Eq. (\ref{eq:Hmu_other}), one finds
\begin{equation}
\label{eq:Hmu_limit}
\overline{H}_\mu \underset{\mu \rightarrow \infty}{\sim} \frac{|\Gamma|}{|\Omega|}\, (\mu/D)^{- \frac12}  \,.
\end{equation}
The right-hand side can be interpreted as the probability that a
uniformly distributed starting point lies within a thin layer of width
$\ell = (\mu/D)^{-1/2}$ near the escape region.  The expression
(\ref{eq:Hmu_limit}) is confirmed on the exactly solvable cases
presented in Appendices \ref{sec:Aexplicit} and \ref{sec:Adisk}.  We
also found a perfect agreement of Eq. (\ref{eq:Hmu_limit}) with two
numerical simulations in which the target is (i) a square concentric
to the confining disk; and (ii) a sphere included within the boundary
of a sphere, both two and three dimensions (results are not shown).

Comparing approximate relations (\ref{eq:Hmu_approx},
\ref{eq:Hmu_limit}), we define the transition death rate $\mu_c$
between two asymptotic regimes:
\begin{equation}
\mu_c = \frac{(|\Omega|/|\Gamma|)^2}{D \overline{\langle \tau_0\rangle}^2} \,.
\end{equation}
Substituting the asymptotic relation (\ref{eq:MFPT_classic}) for the
global MFPT, one finds that $\mu_c \sim D/a^2$ in both two and three
dimensions, where $a$ is the size of the escape region.

\subsection{First passage position}
\label{sec:position}

While we mainly focused on the FPT to the escape region, the
identification and further isolation of the most probable exit
locations is also of practical importance.  In this section, we show
how the distribution of first passage positions (FPP) can be expressed
through the MFPT of mortal walkers.

For this purpose, we recall that the probability flux density,
\begin{equation}
q(t,x; x_0) = - D \frac{\partial}{\partial n} G_t(x;x_0) ,
\end{equation}
obtained from the diffusion propagator $G_t(x; x_0)$, can be
interpreted as the joint probability density for the FPT and the FPP
on the escape region (e.g., see \cite{Caginalp12}).  In particular,
the integral of this function over $x \in \Gamma$ yields the marginal
probability density $\rho(t;x_0)$ for the FPT,
\begin{equation} 
\label{eq:rho_omega}
\rho(t;x_0) = \int\limits_\Gamma dx \, q(t,x;x_0) 
\end{equation}
(see Appendix \ref{sec:A_omega}).  In turn, the integral over time $t$
yields the marginal probability density $\omega_\mu(x;x_0)$ for the
FPP of mortal walkers,
\begin{equation}
\label{eq:omega_mu}
\omega_\mu(x;x_0) = \int\limits_0^\infty dt \, q(t,x;x_0) \, e^{-\mu t} ,
\end{equation}
where we explicitly included the factor $e^{-\mu t}$ to account for
the first-order bulk reaction (see Appendix \ref{sec:A_omega} for
derivation).  Finally, if the starting point $x_0$ is distributed
uniformly, one gets for any $x\in \pa$
\begin{equation}
\label{eq:omega_mu_MFPT}
\begin{split}
\overline{\omega}_\mu(x) & = \frac{1}{|\Omega|} \int\limits_\Omega dx_0 \, \omega_\mu(x;x_0)   
= - \frac{D}{|\Omega|} \int\limits_0^\infty dt \, \frac{\partial}{\partial n} S(t;x) \, e^{-\mu t} \\
& = - \frac{D}{|\Omega|} \frac{\partial \tilde{S}(\mu;x)}{\partial n} = - \frac{D}{|\Omega|} \frac{\partial}{\partial n} \langle \tau_\mu \rangle 
= \frac{D}{\mu |\Omega|} \frac{\partial H_\mu(x)}{\partial n}  . \\
\end{split}
\end{equation}
In other words, the MFPT for mortal walkers also determines their
escape positions.

We illustrate the result (\ref{eq:omega_mu_MFPT}) in the case of an
escape from a ball of radius $R$, for which the MFPT $\langle
\tau_\mu\rangle$ is given in Eq. (\ref{eq:tildeS_ball}).  In the
limit of small death rates, the distribution converges to
$\overline{\omega}_{\mu} \sim 1/\lvert \Gamma \lvert$, where $\lvert
\Gamma \lvert = \lvert \pa \lvert$ is the surface of the
$(d-1)$-sphere, which corresponds to the expected uniform measure over
the sphere.  At large death rates, we find that
$\overline{\omega}_{\mu} \sim \frac{1}{|\Omega|} (\mu/D)^{-1/2}$ in
any dimensions.  The latter relation is expected since it corresponds
to the product of the uniform measure on the sphere and the GEP
defined in Eq. (\ref{eq:Hmu_limit}).

\begin{figure}
\includegraphics[width=80mm]{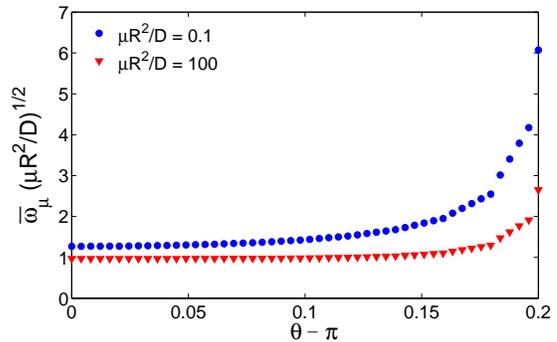} 
\caption{
(Color online) FPP probability density $\overline{\omega}_{\mu}(x)$
(given by Eq. (\ref{eq:omega_disk})), multiplied by $\sqrt{\mu
R^2/D}$, as a function of angular coordinate $\theta$ of the passage
position $x$ on $\Gamma$, for the disk of radius $R$ with an escape
arc $(\pi-\ve,\pi+\ve)$ and $\ve = 0.2$ (see Fig. \ref{fig:scheme}b).
For a small death rate $\mu R^2/D = 0.1$, the density greatly
increases on the edges of the escape region (at $\theta = \ve$).  For
a larger death rate $\mu R^2/D = 100$, the density becomes flatter,
illustrating the convergence to a uniform distribution as $\mu
\to\infty$.  Note that only the half of the density, for $\pi \leq
\theta \leq \pi + \ve$, is shown (the other half being symmetric).}
\label{fig:omega}
\end{figure}

The expression (\ref{eq:omega_mu_MFPT}) is of particular interest in
the study of the narrow escape problem.  As discussed in
\cite{Caginalp12} for immortal walkers, the probability density
$\overline{\omega}_{0}(\theta)$ diverges at the boundaries of the
escape region.  In other words, immortal walkers tend to exit through
the edges of the escape region.  We show that this proclivity is
hindered at large death rates $\mu$, at which
$\overline{\omega}_{\mu}(\theta) \sim (\mu/D)^{-1/2}$ (see
Fig. \ref{fig:omega}).  In the next section, we discuss the
implications of the asymptotic behavior (\ref{eq:omega_mu_MFPT}) at
low and high death rates for the leakage control.

\section{Discussion}
\label{sec:discussion}

\subsection{Leakage control}

In many real-life applications, one needs to fabricate efficient
containers for a safe storage of dangerous species such as nuclear
wastes or toxic chemicals.  For this purpose, one can either improve
the isolation of the container, or incorporate mechanisms to bind,
transform, deactivate or disintegrate dangerous species in the bulk or
on the boundary.  In mathematical terms, the first strategy aims at
shrinking escape regions to increase the MFPT $\langle\tau_0\rangle$
for immortal walkers according to the asymptotic relation
(\ref{eq:MFPT_classic}).  Since a complete isolation is not realistic
(e.g., due to a slow but permanent degradation of container
materials), the exit probability of intact diffusing species (with
$\mu=0$) would be equal to $1$.  In other words, whatever the
isolation improvements are, the leakage is only a matter of time.  It
is therefore important to implement in parallel the second strategy
that aims at reducing the mean lifetime $1/\mu$.  Note that this is a
natural frame for nuclear wastes which disintegrate by a radioactive
decay.

The asymptotic analysis in Sec. \ref{sec:main} has shown how the MFPT
$\langle \tau_\mu\rangle$ and the exit probability $H_\mu(x_0)$ are
affected by the finite lifetime of the walkers.  When the mean
lifetime $1/\mu$ is much larger than the time $\langle
\tau_0\rangle$ needed to find the escape region by immortal walkers,
the decay mechanism weakly affects the escape process.  In particular,
the exit probability remains close to $1$, indicating on a poor
leakage protection.  Improving either the isolation, or the decay
mechanism, or both, one aims at switching to the opposite limit $\mu
\langle \tau_0 \rangle \gg 1$, in which the exit probability can be
significantly reduced.

In this regime, the MFPT $\langle \tau_\mu\rangle$ was shown to
decrease universally as $1/\mu$, i.e., this quantity essentially
reflects the mean lifetime.  In turn, the exit probability
$H_\mu(x_0)$ remains informative and was shown to decay as
stretched-exponential, $\exp(-|x_0-\Gamma|\sqrt{\mu/D})$, when the
death rate $\mu$ increases.  While we could not provide a general form
of the dependence of the exit probability $H_\mu(x_0)$ on the size of
the escape region, we derived the asymptotic behavior of the exit
probability for several specific domains (see Appendices
\ref{sec:Aexplicit} and \ref{sec:Adisk}).  

The leakage of many species uniformly distributed in a container can
be characterized by the global exit probability $\overline{H}_\mu$
which exhibits a much slower decay with $\mu$, as suggested by
approximations (\ref{eq:Hmu_approx}, \ref{eq:Hmu_limit}).  The
approximation (\ref{eq:Hmu_approx}) can be interpreted as the result
of competition of two first-order kinetics:
\begin{align}
\mathrm{Dead} \underset{\mu}{\longleftarrow} \mathrm{Alive} \underset{k_e}{\longrightarrow} \mathrm{Exited}
\end{align}
where $k_e = 1/\overline{\langle \tau_0 \rangle}$ is the exit rate in
absence of a ``death mechanism''.  In the stationary state, the law of
mass action \cite{Krapivsky} predicts that the proportion of dead
$n_D$ to exited $n_E = 1 - n_D$ walkers reads $n_D/n_E = \mu \,
\overline{\langle \tau_0 \rangle}$, from which $n_E= 1/(1+\mu \,
\overline{\langle \tau_0 \rangle})$.  Since the proportion $n_E$ of
exited walkers is precisely the exit probability $\overline{H}_\mu$,
one recovers the approximation (\ref{eq:Hmu_approx}).  One can
conclude that the GEP remains close to $1$ at small $\mu$ even for
high quality isolation (small $\ve$), whereas shrinking the escape
region allows one to significantly improve the safety of a container
when $\mu R^2/D \gtrsim 1$.  At very large death rates, the above
kinetic argument fails, and the GEP is determined by the universal
asymptotic relation (\ref{eq:Hmu_limit}).

Note also that active protecting mechanisms can be implemented not
only in the bulk, but also on the container boundary, including the
escape regions.  As shown in \cite{Grebenkov16}, the introduction of
an energetic or entropic barrier at the escape region significantly
affects the MFPT $\langle \tau_0\rangle$ for immortal walkers, making
the escape process ``barrier-limited'' instead of
``diffusion-limited''.  In particular, the conventional asymptotic
behavior (\ref{eq:MFPT_classic}) is replaced by a faster divergence as
$(\kappa/D)\, \epsilon^{1-d}$, where $\kappa$ is the barrier
reactivity or permeability.  In other words, the presence of energetic
or entropic barrier greatly improves the quality of isolation and
allows one to reduce the exit probability.  In the same vein,
long-ranged repulsive interactions keep diffusing species expelled
from the boundary and thus increase the MFPT of immortal walkers
\cite{Grebenkov16}.  In addition, our general result
(\ref{eq:omega_mu_MFPT}) concerning the distribution of exit positions
can be used to design protecting shields that would be adapted to the
death rate of the toxic reactant.  In particular, we showed that
highly mortal walkers hit uniformly the escape region, in contrast to
immortal walkers that mainly exit through the edges of the escape
region (see discussion in Sec.
\ref{sec:position}).

\subsection{Scaling argument for mRNA translation}

As another application of our results, we propose a scaling argument
to express the variability of the mRNA lifetime between E. Coli, Yeast
and Human cells in terms of the cell volume and the number of
ribosomes.  In this biological context, the escape region is the
surface of ribosomes, while the exit probability can be interpreted as
the probability for an mRNA to encounter a ribosome before being
degradated.  Assuming that the asymptotic relation
(\ref{eq:Hmu_limit}) holds (i.e., that mRNA translation occurs in the
high degradation rate regime), we get the scaling relation
\begin{align} 
\label{eq:scaling_argument}
\frac{N s_r}{V} \, \sqrt{D/\mu} \sim \overline{H}_\mu 
\end{align}
between the degradation rate $\mu$, the cell volume $V$, and the total
ribosome surface $S = N s_r$ that mRNA should hit in order to trigger
a translation, where $s_r$ is the typical reactive surface of one
ribosome and $N$ is the number of ribosomes in the cell.

In addition, we assume that (i) the diffusion coefficient $D$ is the
same for different species \cite{Milo2015} and (ii) that an identical
``success rate'' $\overline{H}_\mu$ in the reaction kinetics should be
maintained for all cells to ensure an efficient translation mechanism.
From Eq. (\ref{eq:scaling_argument}), the ratio $N/(V \sqrt{\mu})$
should then be approximatively constant among different species.  This
scaling argument is confirmed by comparing E. Coli, Yeast, and Human
cells (Table \ref{tab:cells}).  Using the known data for one cell
type, one should thus be able to estimate the degradation rate of mRNA
in another cell type from its volume and the number of ribosomes.
Similar arguments could hold for the transcription problem, involving
the search of a specific DNA site by transcription factors.  As
discussed in \cite{Godec16}, it appears that only the fastest $0.01\%$
to $1\%$ of transcription factors actually matter for the cellular
response.

In summary, our results for the GEP provides a new interpretation for
the cell size scaling problem, which has received a large attention
within the recent years (see \cite{Pan14} and references therein).  By
narrowing the spread of the first passage times, the degradation rate
might be used to focus the temporal cellular response to external
perturbations.

\begin{table}
\begin{center}
\begin{tabular}{| c | c | c | c |} \hline
                   & E. Coli & Yeast  & Human \\  \hline
$N$                & $10^4$  & $10^5$ & $10^6$ \\ 
$V$ (in $\mu$m$^3$) &    $2$  &  $40$  &  $2000$ \\ 
$1/\mu$ (in min)    &    $5$  &  $20$  &  $600$  \\  \hline
$10^{-4}~N/(V\sqrt{\mu})$ & \multirow{2}{*}{$1.1$} & \multirow{2}{*}{$1.1$} & \multirow{2}{*}{$1.2$} \\ 
(in $\mu$m$^{-3}\cdot$min$^{1/2}$) & & & \\  \hline
\end{tabular}
\end{center}
\caption{
The number of ribosomes ($N$), the cell volume ($V$), and the mRNA
degradation rate ($\mu$) for E. Coli, Yeast, and Human fibroblast
cells \cite{Milo2015}.  }
\label{tab:cells}
\end{table}

\subsection{Extensions}

The escape problem for mortal walkers can be extended in various ways.
First, the diffusion equation in Eqs. (\ref{eq:propagator}) can be
replaced by a more general backward Fokker-Planck (or Kolmogorov)
equation to account for the effect of external forces or potentials
\cite{Gardiner}.  For instance, external forces can model the effect
of chemotactic gradient which was proved to be relevant for the egg
search problem by spermatozoa \cite{Alvarez14}.  Second, the Dirichlet
boundary condition on the escape region can be replaced by a Robin
condition to model the presence of a recognition step, partial
reflections, stochastic gating, or microscopically heterogeneous
distribution of exit channels
\cite{Collins49,Sano79,Sapoval94,Benichou00,Grebenkov06,Singer08,Bressloff08,Grebenkov15}.
The survival probability and the distribution of FPTs to the whole
partially absorbing boundary have been earlier studied for immortal
walkers \cite{Grebenkov10a,Grebenkov10b,Rojo12}.  Third, one can
consider intermittent processes, with alternating phases of bulk and
surface diffusion
\cite{Benichou10,Benichou11,Rojo11,Rupprecht12a,Rupprecht12b,Rojo13}.
Fourth, one can investigate the effect of heterogeneous materials or
multiple layers of a container with different diffusion or trapping
properties (this is a typical situation for nuclear waste storage).
Fifth, the retarding effect of geometric or energetic traps can be
included by considering the continuous time random walk (CTRW) with a
fat-tailed waiting time distribution
\cite{Klafter,Bouchaud90,Metzler14b,Yuste07,Grebenkov10a,Grebenkov10b}.
The extension of these problems to mortal walkers consists in
replacing the Laplace transform variable $p$ by $p + \mu$, as in
Eq. (\ref{eq:Smu}) for normal diffusion.  

To illustrate this point, we consider an extension to CTRW for which
the propagator obeys the fractional diffusion equation, while the
Laplace-transformed survival probability satisfies the Helmholtz
equation
\begin{equation}
\bigl[p^{1-\alpha} D_\alpha \Delta - p\bigr] \tilde{S}^\alpha(p;x_0) = -1 ,
\end{equation}
with the same boundary conditions as for normal diffusion, where
$0<\alpha<1 $ is the scaling exponent, and $D_\alpha$ is the
generalized diffusion coefficient (in units m$^2$/s$^\alpha$).
Changing the variable $p$ to $D p^\alpha/D_\alpha$, the solution of
this equation can be expressed through the earlier obtained
$\tilde{S}(p;x_0)$:
\begin{equation}
\tilde{S}^{\alpha}(p;x_0) = \frac{p^{\alpha-1} D}{D_\alpha} \, \tilde{S}( Dp^\alpha/D_\alpha ;x_0) .
\end{equation}
From this relation, one immediately retrieves that the MFPT of
immortal walkers is infinite: $\langle \tau_{0,\alpha}\rangle =
\tilde{S}^{\alpha}(0;x_0) = \infty$ because the walkers can be trapped
for long periods of time until they reach the escape region.  In
contrast, the MFPT for mortal walkers is finite: 
\begin{equation}
\langle \tau_{\mu,\alpha}\rangle = \frac{\mu^{\alpha-1} D}{D_\alpha}\, \tilde{S}(D\mu^\alpha/D_\alpha ;x_0)
= \frac{\mu^{\alpha-1} D}{D_\alpha}\, \langle \tau_{D\mu^\alpha/D_\alpha} \rangle .
\end{equation}
In fact, too long trajectories whose contribution led to divergence of
the MFPT for CTRW, are eliminated because of a finite lifetime of the
walker.  In other words, the MFPT for mortal CTRW is related to the
MFPT for mortal normal walkers with a modified death rate: $\mu_\alpha
= D\mu^\alpha/D_\alpha$.  Finally, the exit probability for mortal
CTRWs that can be defined in analogy with Eq. (\ref{eq:Hexit}) as
\begin{equation}
H_{\mu,\alpha}(x_0) = \int\limits_0^\infty dt \, e^{-\mu t} \, \left(-\frac{\partial S^\alpha(t; x_0)}{\partial t}\right) ,
\end{equation}
is simply
\begin{equation}
H_{\mu,\alpha}(x_0) = H_{\mu_\alpha}(x_0) .
\end{equation}
One can therefore apply the results from Sec. \ref{sec:main} to mortal
CTRWs.

In all these extensions, the first-order bulk kinetics or,
equivalently, an exponentially distributed lifetime of the walker,
controls the duration of trajectories, assigning smaller weights to
longer trajectories.  We also mention the possibility of considering
other lifetime distributions $Q(t)$ beyond the exponential one.  In
this case, Eq. (\ref{eq:Smu0}) is still applicable but one needs to
get the survival probability in time domain by inverse Laplace
transform.  Finally, if the death mechanism is coupled to diffusion
(e.g., in the case of a space-dependent death rate), one has to treat
the whole diffusion-reaction mixed boundary value problem.

\section{Conclusion}
\label{sec:conclusion}

We formulated the escape problem to mortal walkers and investigated
how their finite lifetime drastically affects the survival and exit
probabilities.  The latter is the likelihood of escape or leakage from
the confining container and can thus characterize its isolation
quality.  We focused on the most relevant case of a first-order bulk
kinetics or, equivalently, an exponentially distributed lifetime, for
which the problem is reduced to finding the Laplace-transformed
survival probability for immortal walkers.  We derived the upper and
lower bounds for the MFPT $\langle \tau_\mu\rangle$ and the exit
probability $H_\mu(x_0)$ and analyzed their asymptotic behavior at
small and large death (or reaction) rates.  When the mean lifetime,
$1/\mu$, is much larger than the MFPT for immortal walkers, $\langle
\tau_0\rangle$, the exit probability remains close to $1$, meaning a
poor isolation.  In this situation, the leakage of dangerous species
is just a matter of time.  For a safer protection, one needs both to
improve the boundary isolation by shrinking escape regions (thus
increasing $\langle \tau_0\rangle$), and to implement efficient
trapping, binding or deactivation mechanisms (thus increasing $\mu$).
Improving only one of these aspects is not sufficient to significantly
reduce $H_\mu(x_0)$.  For the volume-averaged, or global exit
probability $\overline{H}_\mu$ that quantifies the overall safety of a
container, we obtained two approximations at intermediate and large
death rates.  The quality of the obtained analytical results for
general confining domains was confirmed by comparison with several
explicitly solvable cases, and with numerical simulations.  We also
introduced and investigated the distribution of the first passage
positions as a mathematical ground for design and optimization of
containers.  The density $\overline{\omega}_\mu(x)$ was shown to
exhibit a transition from a singular function highly localized near
edges of the escape region for immortal walkers, to asymptotically
uniform density at large death rates.  Various extensions and
applications of the studied escape problem have been discussed
including the leakage control of dangerous chemicals and scaling
relation for mRNA translation mechanism.

\begin{acknowledgments}
The authors acknowledge the support under Grant
No. ANR-13-JSV5-0006-01 of the French National Research Agency.  We
thank T. Saunders and M. Howard for interesting discussions about the
cell size regulation problem.
\end{acknowledgments}

\appendix
\section{Explicit results for rotation-invariant domains}
\label{sec:Aexplicit}

\subsection{Escape from a ball}
\label{sec:Aball}

The escape through the whole boundary of a ball is the simplest and
the most studied case.  The separation of variable allows one to
derive an explicit spectral representation of the propagator in terms
of Laplacian eigenfunctions, from which other quantities are deduced
\cite{Redner,Gardiner}.  In particular, the Laplace-transformed
survival probability from the ball $B_d(R)$ of radius $R$ in $\R^d$
reads (e.g., see \cite{Yuste07,Grebenkov10b}):
\begin{equation}
\label{eq:tildeS_ball}
\tilde{S}_{B_d(R)}(p;x_0) = \frac{1}{p}\left(1 - \left(\frac{|x_0|}{R}\right)^{1-\frac{d}{2}}
\frac{I_{\frac{d}{2}-1}(|x_0| \sqrt{p/D})}{I_{\frac{d}{2}-1}(R\sqrt{p/D})}\right) ,
\end{equation}
where $I_\nu(z)$ is the modified Bessel function of the first kind,
and $x_0$ is the starting point.  From this expression, one can
immediately deduce the MFPT, the exit probability and the GEP for
mortal walkers.  The asymptotic behavior at small and large $p$ can be
easily obtained.  

In one dimension, this is the escape problem from an interval $(-R,R)$
with two escape points $\pm R$ or, equivalently, from an interval
$(0,R)$ with one escape point at $x = R$ and reflecting endpoint at $x
= 0$.  In this case, Eq. (\ref{eq:tildeS_ball}) reduces to
\begin{align}
\left\langle \tau_\mu \right\rangle = \frac{1}{\mu} \left(1 - \frac{\cosh\bigl(x_0\sqrt{\mu/D}\bigr)}{\cosh \bigl(R\sqrt{\mu/D}\bigr)} \right) ,
\end{align}
from which the exit probability and the GEP read as
\begin{align}
H_{\mu}(x_0) = \frac{\cosh\bigl(x_0 \sqrt{\mu/D}\bigr)}{\cosh \bigl(R\sqrt{\mu/D}\bigr)} 
\end{align}
and
\begin{align}
\overline{H}_{\mu} = \frac{\tanh \bigl(R \sqrt{\mu/D}\bigr)}{R \sqrt{\mu/D}} \underset{\mu \rightarrow \infty}{\sim}  \frac{1}{R \sqrt{\mu/D}}  \,.
\end{align}

In two dimensions, one gets the MFPT and global MFPT
\begin{align} 
\label{eq:uniform_mfpt_2D}
\left\langle \tau_\mu \right\rangle =  \frac{1}{\mu} \left( 1-\frac{I_0\bigl(r\sqrt{\mu/D}\bigr)}{I_0\bigl(R \sqrt{\mu/D}\bigr)}\right)
\end{align}
and 
\begin{align} 
\label{eq:uniform_gmfpt_2D}
\overline{\langle \tau_\mu \rangle} = \frac{1}{\mu} \left(1 - \frac{2 I_1\bigl(R\sqrt{\mu/D}\bigr)}{R \sqrt{\mu/D}\, I_0\bigl(R \sqrt{\mu/D}\bigr)}  \right) ,
\end{align}
where $r = |x_0|$.  The exit probability and the GEP read
\begin{align} 
\label{eq:uniform_exit_2D}
H_{\mu}(r) = \frac{I_0\bigl(r\sqrt{\mu/D}\bigr)}{I_0\bigl(R \sqrt{\mu/D}\bigr)} 
\underset{\mu \rightarrow \infty}{\sim}  \frac{e^{-(R-r)\sqrt{\mu/D}}}{\sqrt{r/R}}
\end{align}
and
\begin{align} 
\label{eq:uniform_geq_2D}
\overline{H}_{\mu} = \frac{2 I_1\bigl(R \sqrt{\mu/D}\bigr)}{R \sqrt{\mu/D} \, I_0\bigl(R \sqrt{\mu/D}\bigr)} 
\underset{\mu \rightarrow \infty}{\sim}  \frac{2}{R \sqrt{\mu/D}} \,.
\end{align}
This asymptotic behavior agrees with Eq. (\ref{eq:Hmu_limit}).

In three dimensions, the expressions for the MFPT and global MFPT are
\begin{align} 
\label{eq:uniform_mfpt_3D}
\left\langle \tau_\mu \right\rangle = \frac{R e^{(R-r)\sqrt{\mu/D}} -R e^{(R+r)\sqrt{\mu/D}} + r e^{2 R \sqrt{\mu/D}} - r}
{\mu r \bigl(e^{2 R \sqrt{\mu/D}}-1\bigr)}
\end{align}
and
\begin{align} 
\label{eq:uniform_gmfpt_3D}
\overline{\langle\tau_\mu \rangle} = \frac{1}{\mu} \left(1 - \frac{3 R \sqrt{\mu/D} \, \coth \bigl(R\sqrt{\mu/D}\bigr)+3}{R^2 \mu/D} \right) .
\end{align}
The exit probability reads
\begin{align} 
\label{eq:uniform_ep_3D}
H_{\mu}(r) = \frac{R \sinh \bigl(r\sqrt{\mu/D}\bigr)}{r \sinh\bigl(R \sqrt{\mu/D}\bigr)} 
 \underset{\mu \rightarrow \infty}{\sim}  \frac{R}{r} e^{-(R-r)\sqrt{\mu/D}} \,,
\end{align}
where the limit is taken for starting points $x_0$ sufficiently far
away from the boundary.  The GEP reads
\begin{align} 
\label{eq:uniform_gep_3D}
\overline{H}_{\mu} = 3 \frac{R \sqrt{\mu/D} \coth \bigl(R \sqrt{\mu/D}\bigr)-1 }{R^2 \mu/D} 
\underset{\mu \rightarrow \infty}{\sim} \frac{3}{R\sqrt{\mu/D}} \,,
\end{align}
in agreement with Eq. (\ref{eq:Hmu_limit}).

\subsection{Concentric escape region}

For most applications, however, the escape region presents only a
small part of the boundary.  We consider another example of
rotation-invariant domains with two concentric boundaries, $\{
x\in\R^d ~:~ a < |x| < R\}$, with the escape region at the inner
boundary of radius $a$, while the outer boundary of radius $R$ is
fully reflecting.  The rotation invariance of the domain leads to
explicit representations of the Laplace-transformed survival
probability, the exit probability and the GEP.  In particular,
Eq. (\ref{eq:tildeS}) becomes
\begin{align}
\label{eq:Sradial}
\left(\frac{\partial^2}{\partial r^2} + \frac{d-1}{r} \frac{\partial }{\partial r} - \frac{p}{D} \right) \tilde{S}(p; r) = - \frac{1}{D} \,
\end{align}
subject to two boundary conditions:
\begin{equation}
\tilde{S}(p; a) = 0, \qquad  \left(\frac{\partial}{\partial r} \tilde{S}(p; r)\right)_{r=R} = 0.
\end{equation}
In two dimensions, the solution of this equation is
\begin{align}
\label{eq:S_concentric_2d}
\tilde{S}(p; r) = \frac{1}{p} \left[1 - \frac{I_1(\sqrt{s}) K_0\bigl(\sqrt{s} \frac{r}{R}\bigr) + K_1(\sqrt{s}) I_0\bigl(\sqrt{s} \frac{r}{R}\bigr)}
{I_1(\sqrt{s}) K_0\bigl(\sqrt{s} \frac{a}{R} \bigr) + K_1(\sqrt{s}) I_0\bigl(\sqrt{s}\frac{a}{R}\bigr)} \right] ,
\end{align}
where $s = pR^2/D$ and $K_\nu(z)$ is the modified Bessel function of
the second kind.  From this expression, one gets
\begin{align}
\label{eq:H_concentric_2d}
H_\mu(r) = \frac{I_1(\sqrt{s}) K_0\bigl(\sqrt{s} \frac{r}{R}\bigr) + K_1(\sqrt{s}) I_0\bigl(\sqrt{s} \frac{r}{R}\bigr)}
{I_1(\sqrt{s}) K_0\bigl(\sqrt{s} \frac{a}{R} \bigr) + K_1(\sqrt{s}) I_0\bigl(\sqrt{s}\frac{a}{R}\bigr)} \,,
\end{align}
with $s = \mu R^2/D$, while the GEP reads
\begin{align}  
\label{eq:geq_2D}
\overline{H}_\mu = \frac{2 \frac{a}{R} \, s^{-1/2} \bigl[I_1(\sqrt{s}) K_1\bigl(\sqrt{s}\frac{a}{R}\bigr) - K_1(\sqrt{s}) I_1\bigl(\sqrt{s}\frac{a}{R}\bigr)\bigr]}
{\bigl(1-(\frac{a}{R})^2\bigr)\bigl[I_1(\sqrt{s}) K_0\bigl(\sqrt{s} \frac{a}{R} \bigr) + K_1(\sqrt{s}) I_0\bigl(\sqrt{s}\frac{a}{R}\bigr)\bigr]} \,.
\end{align}
Note that the MFPT for immortal walkers is
\begin{equation}
\langle \tau_0 \rangle = \tilde{S}(0;r) = \frac{R^2 \ln(r/a)}{2D} - \frac{r^2 - a^2}{4D} \,,
\end{equation}
that diverges as $\ln(1/a)$ for $a\to 0$, as expected.

In three dimensions, the solution of Eq. (\ref{eq:Sradial}) is
\begin{align}
\label{eq:S_concentric_3d}
\tilde{S}(p; r) = \frac{1}{p} \left[1 - \frac{a}{r}  e^{-\sqrt{s} (r-a)/R} 
\frac{\frac{\sqrt{s}-1}{\sqrt{s}+1} + e^{-2\sqrt{s} (1-r/R)}}{\frac{\sqrt{s}-1}{\sqrt{s}+1} +  e^{-2\sqrt{s}(1-a/R)}}  \right],
\end{align}
from which
\begin{align}
\label{eq:H_concentric_3d}
H_\mu(r) = \frac{a}{r}  e^{-\sqrt{s} (r-a)/R} \frac{\frac{\sqrt{s}-1}{\sqrt{s}+1} + e^{-2\sqrt{s} (1-r/R)}}
{\frac{\sqrt{s}-1}{\sqrt{s}+1} +  e^{-2\sqrt{s}(1-a/R)}} \,,
\end{align}
and
\begin{align} 
\label{eq:geq_3D}
\overline{H}_\mu = \frac{3 \frac{a}{R} \left( \frac{\sqrt{s}-1}{\sqrt{s}+1} \bigl(1 + \frac{a}{R} \sqrt{s}\bigr) 
+ e^{-2\sqrt{s}(1-a/R)} \bigl(1 - \frac{a}{R} \sqrt{s}\bigr)\right)}
{s \bigl(1-\frac{a^3}{R^3}\bigr) \left( \frac{\sqrt{s}-1}{\sqrt{s}+1} +  e^{-2\sqrt{s}(1-a/R)} \right)} \,.
\end{align}
Note that the MFPT for immortal walkers is
\begin{equation}
\langle \tau_0 \rangle = \tilde{S}(0;r) = \frac{R^3(r-a)}{3Dra} - \frac{r^2 - a^2}{6D} \,,
\end{equation}
that diverges as $1/a$ for $a\to 0$, as expected.

The explicit form of the exit probability, $H_\mu(x_0)$, and the GEP,
$\overline{H}_\mu$, helps us to investigate three asymptotic regimes
of weak ($\mu \ll D/R^2$), intermediate ($D/R^2 \ll \mu \ll D/a^2$),
and strong ($\mu \gg D/a^2$) death rates.  Note that the second regime
emerges only under the narrow escape condition $a \ll R$ that we
assume here.  We start with the three-dimensional case for which the
analysis is much simpler.

In the weak death rate regime, the Taylor expansion
(\ref{eq:Hmu_small}) reduces the analysis to computing the moment of
$\tau_0$ which are given by Eq. (\ref{eq:tau_moments}) and can be
found explicitly from the exact formula (\ref{eq:H_concentric_3d}).
When $\mu \gg D/R^2$, the GEP from Eq. (\ref{eq:geq_3D}) is greatly
simplified to
\begin{align} 
\overline{H}_\mu \simeq \frac{3a  \bigl(1 + a\sqrt{\mu/D}\bigr)}{R^3 \mu/D} \,,
\end{align}
from which the intermediate and strong death rate regimes can be
distinguished:
\begin{align} 
\label{eq:geq_3D_asymptotic}
\overline{H}_\mu \simeq \frac{3a}{R^3} \times \left\{ \begin{array}{l l}  D/\mu & (D/R^2 \ll \mu \ll D/a^2), \\
a \sqrt{D/\mu} & \hskip 15mm (\mu \gg D/a^2). \\ \end{array} \right.
\end{align}
The latter asymptotic regime agrees with Eq. (\ref{eq:Hmu_limit}).  In
both regimes, the exit probability is simplified when $r$ is not too
close to $R$
\begin{align} \label{eq:exit_3D_asymptotic}
H_\mu(r) \simeq \frac{a}{r} \exp\bigl(-(r-a)\sqrt{\mu/D} \bigr) \,.
\end{align}

In two dimensions, one gets 
\begin{align}  \label{eq:geq_2D_intermediate}
\overline{H}_\mu \sim \frac{2}{R^2 \mu/D \bigl(-\gamma - \ln \frac{a\sqrt{\mu/D}}{2}\bigr)} 
\end{align}
for intermediate death rates, and
\begin{align}  \label{eq:geq_2D_asymptotic}
\overline{H}_\mu \underset{\mu \rightarrow \infty}{\sim} \frac{2 a }{R^2 \sqrt{\mu/D}} 
\end{align}
in the limit of strong death rates ($\mu \gg D/a^2$), where $\gamma$
is the Euler constant.  Again, this result agrees with the general
relation (\ref{eq:Hmu_limit}).  In the limit $\mu \gg D/R^2$ (or $s
\gg 1$), one gets
\begin{align}  \label{eq:exit_2D}
H_\mu(r) \underset{\mu \rightarrow \infty}{\sim} \frac{\sqrt{\pi}\,  e^{-r \sqrt{\mu/D}} \bigl(1 + e^{-2(R-r)\sqrt{\mu/D}}\bigr)}
{\sqrt{2}\, (r^2\mu/D)^{1/4} \bigl(\gamma + \ln \frac{a \sqrt{\mu/D}}{2} \bigr)} \,.
\end{align}

\section{Exit time from a ball with reflecting obstacles}
\label{sec:conjecture}

\begin{figure}
\begin{center}
\includegraphics*[width=28mm]{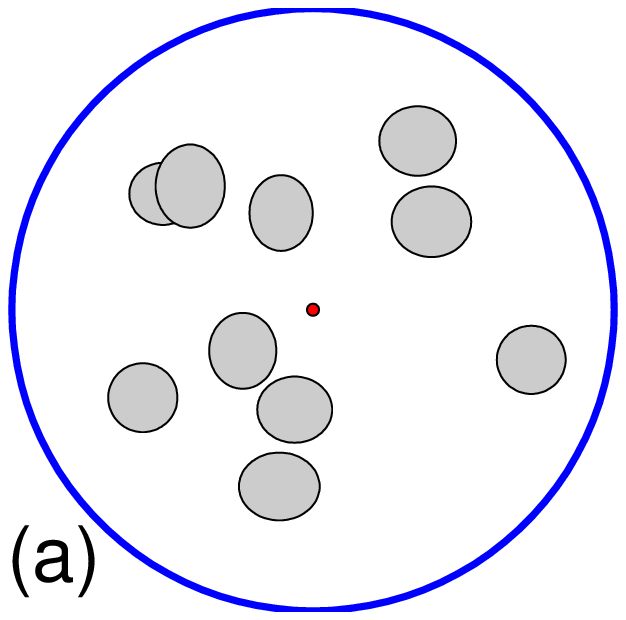} 
\includegraphics*[width=28mm]{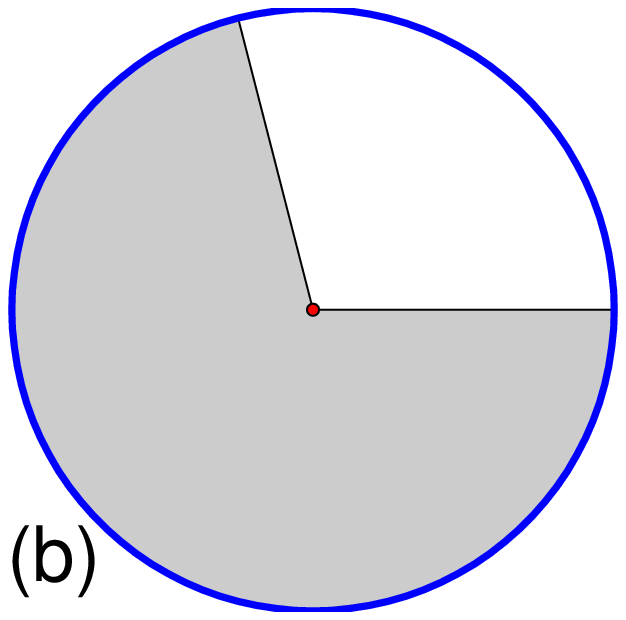} 
\includegraphics*[width=28mm]{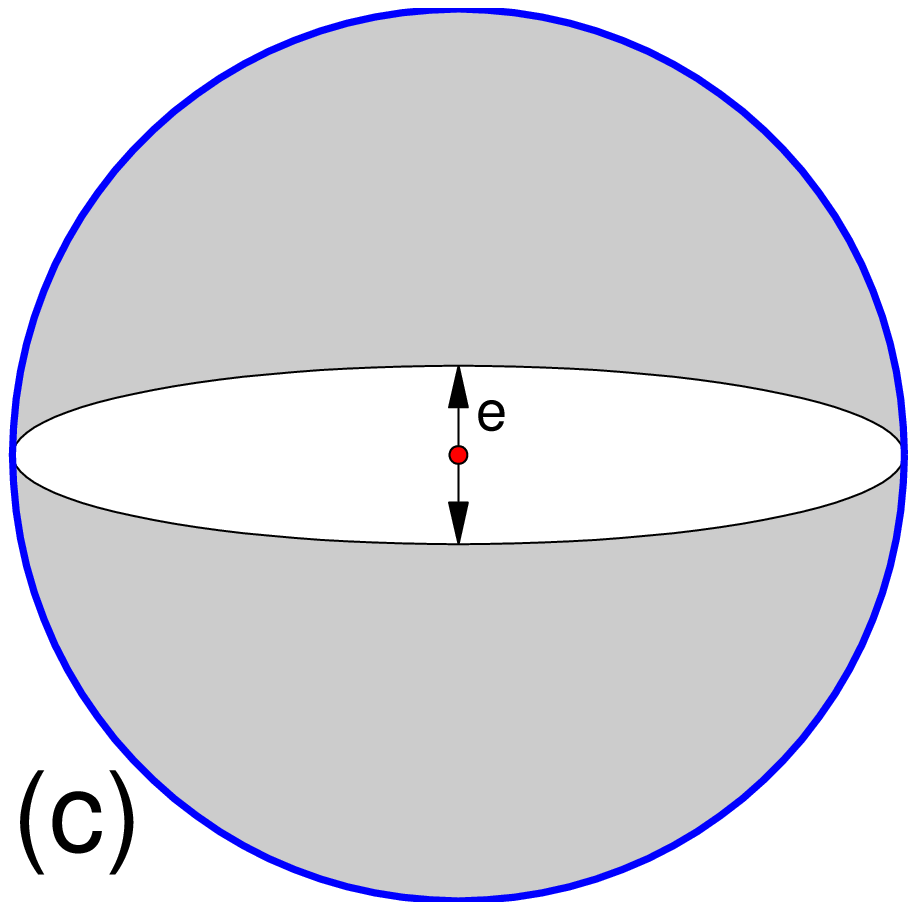} 
\end{center}
\caption{
(Color online) Illustrations of the escape problem from a ball with
reflecting obstacles (gray shadowed regions). }
\label{fig:disk}
\end{figure}

The lower bound of the Laplace-transformed survival probability relies
on the inequality (\ref{eq:Sineq}) which states that the FPT from a
point $x_0$ to the escape region $\Gamma$ is greater than the FPT from
the origin of the ball $B = B_d(|x_0-\Gamma|)$ of radius
$|x_0-\Gamma|$ to its boundary.  This property is based on two
statements: (i) the FPT to the escape region is greater than the first
exit time from the subdomain $\Omega \cap B$ through the boundary of
the ball $B$, and (ii) the latter is equal to or greater than the FPT
from the origin of the ball $B$ to its boundary.  The first statement
follows from the continuity of Brownian motion: before reaching
$\Gamma$, a particle must cross the boundary of the ball $B$.
However, the second statement is less evident.  It claims that the
presence of reflecting obstacles inside a ball (here, the boundary
$\pa \cap B$) cannot speed up the exit from this ball
(Fig. \ref{fig:disk}a).  Qualitatively, this statement sounds natural:
reflecting obstacles ``shield'' some boundary points from the center
and thus potentially increase the FPT.  Note also that the presence of
reflecting obstacles may not affect the FPT at all, as illustrated by
the case of a sector, for which the FPT is exactly the same as for the
whole disk (Fig. \ref{fig:disk}b).

We mention that this claim cannot be generalized to any starting point
or any domain.  Indeed, if the starting point was not at the origin of
the ball, adding a concentric spherical reflecting obstacle of radius
$r = |x_0|$ would certainly speed up the exit.  For instance, for an
annulus $r < |x| < R$ with reflecting inner circle at $r$, the MFPT
from the starting point $(r,\theta)$ (in polar coordinates) to the
outer circle at $R$ is
\begin{equation}
\langle \tau_0 \rangle = \frac{R^2 - r^2}{4D} + \frac{r^2}{2D} \, \ln(r/R).  
\end{equation}
The first term, which is the MFPT for the disk without obstacles, is
decreased by the second negative term.

Without having a rigorous mathematical proof for this statement, we
illustrate it on several examples.  First, let us consider a case when
the reflecting obstacles covers most of the volume of the ball
$B_d(R)$ (see Fig. \ref{fig:disk}c).  When the width $e$ of the
remaining domain accessible to diffusion shrinks to zero, the search
process becomes $(d-1)$-dimensional (e.g., for $d=2$, the walkers
diffuse along one-dimensional interval $(-R,R)$ and search for its
endpoints).  We expect that the survival probability at any given time
$t$ increases as $e$ decreases.  This behavior is exemplified by
considering the limiting case $e=0$: the survival probability
$S_{B_{d-1}(R)}(t;0)$ (i.e., when the escape region $\Gamma$ is the
boundary of $(d-1)$-dimensional ball $B_{d-1}(R)$) is larger than the
survival probability $S_{B_d(R)}(t;0)$ that corresponds to the escape
from the original ball $B_d(R)$.  This follows from the fact that
$\tilde{S}_{B_d(R)}(\mu;0)$ from Eq. (\ref{eq:tildeS_ball0}) is a
decreasing function of the dimensionality $d$ for all $\mu \geq 0$.
In other words, the Laplace-transformed survival probability from the
center of the ball $B_d(R)$ is smaller than its counterpart in the
lower dimension $d-1$, i.e., $S_{B_{d-1}(R)}(\mu;0)$.  One can see
that drastic volume occupancy by reflecting obstacles increases the
FPT.  Note also that the MFPT from the center of a ball without
obstacles, $R^2/(2dD)$, also increases with the dimensionality
reduction.

\begin{figure}
\includegraphics[width=70mm]{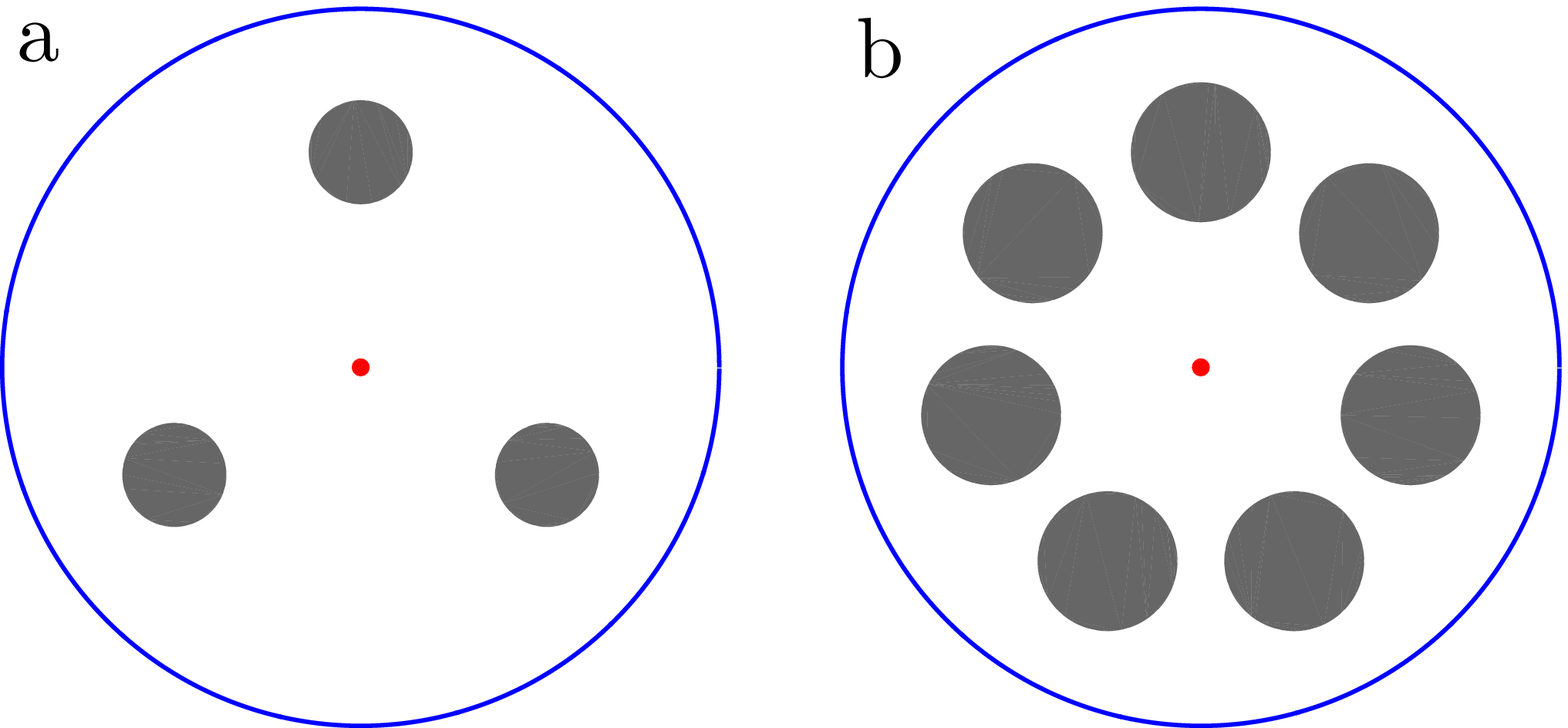} 
\includegraphics[width=80mm]{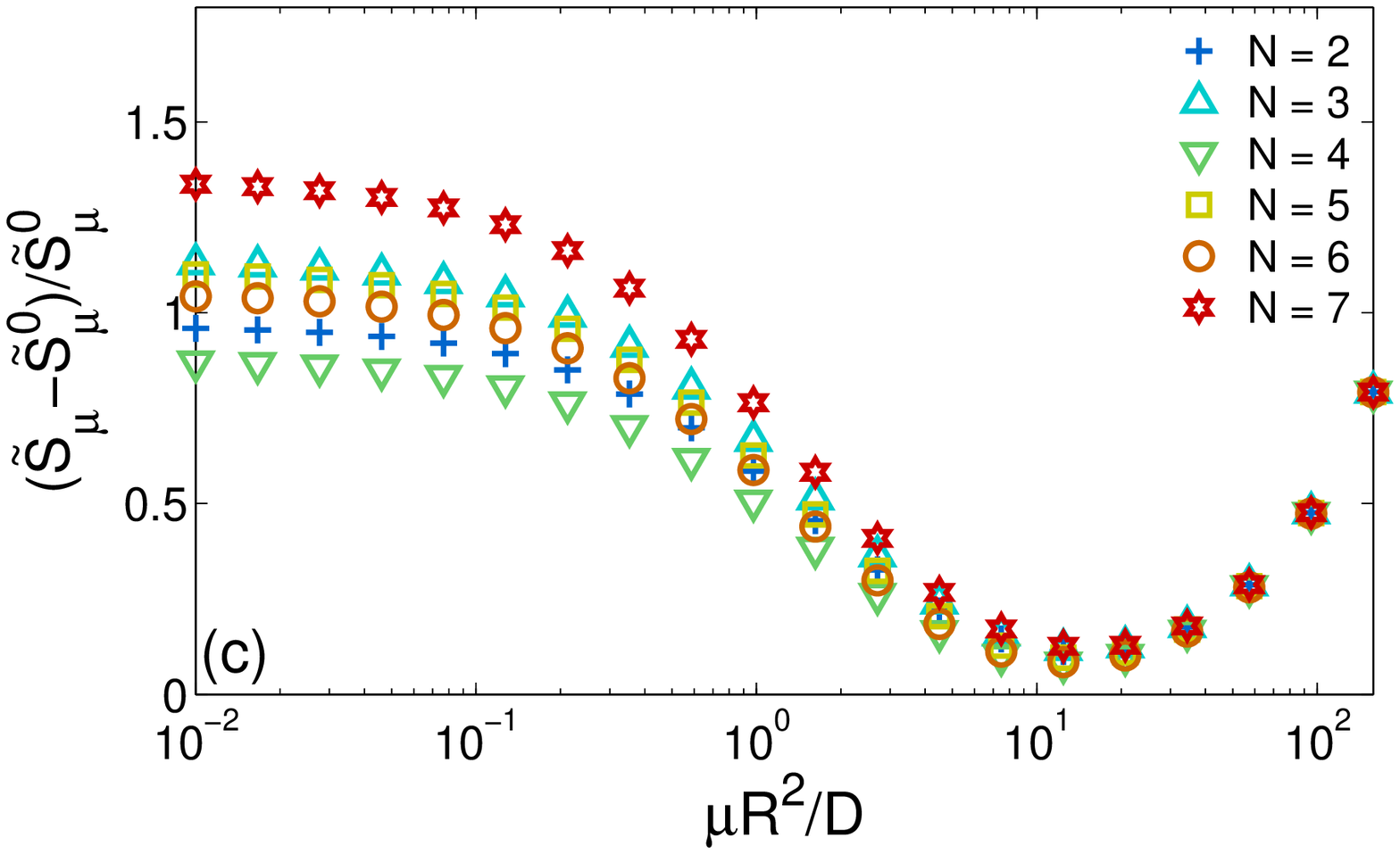} 
\includegraphics[width=80mm]{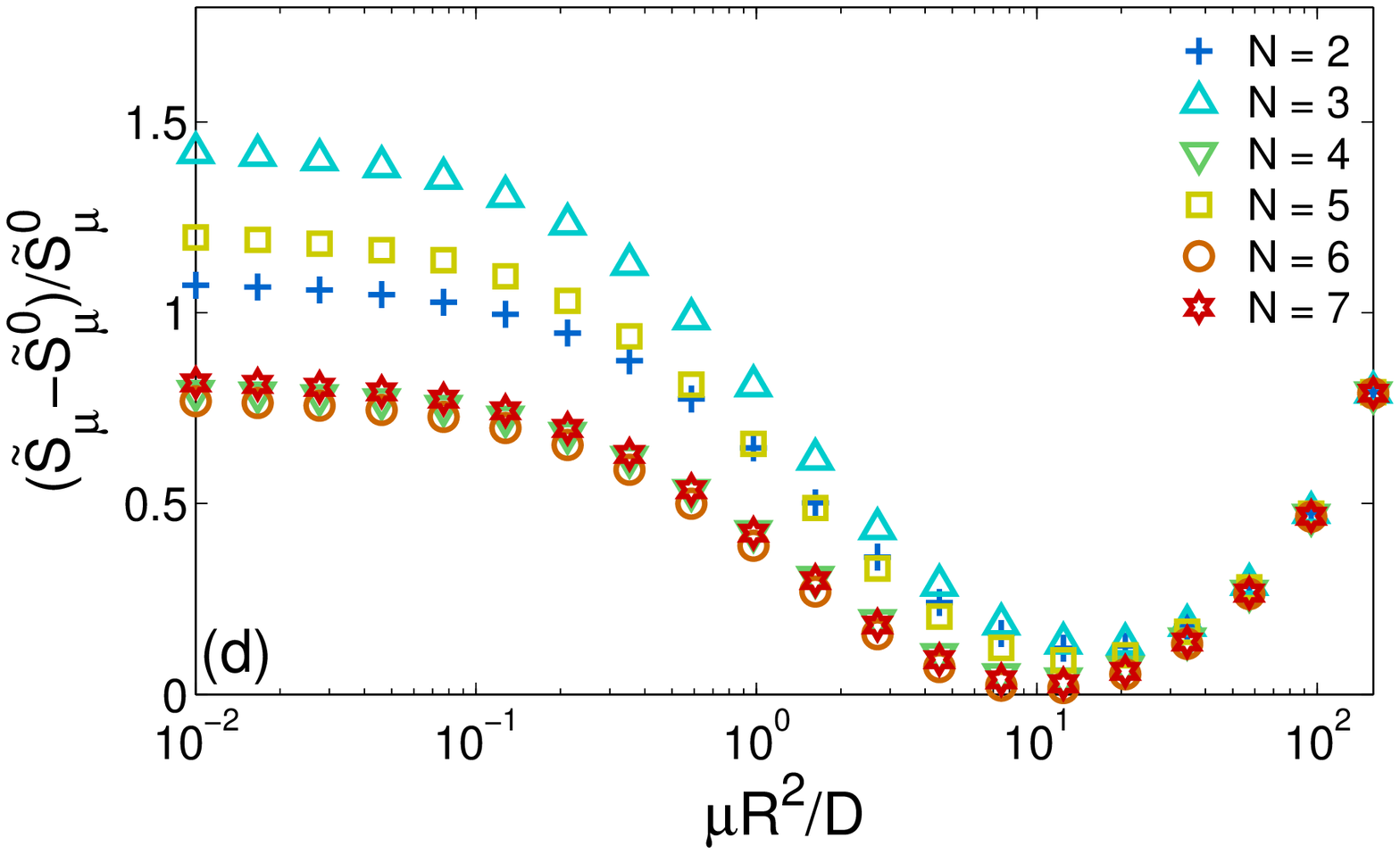} 
\caption{
(Color online) {\bf (a,b)} Examples of regular arrangments of
reflecting circular obstacles (gray shadowed regions) of radius $r$
inside the unit disk, with $r = 0.1$ {\bf (a)} and $r = 0.2$ {\bf
(b)}.  {\bf (c-d)} Relative difference between the Laplace-transformed
survival probability in the presence of obstacles, $\tilde{S}_{\mu}$
(computed numerically), and that without obstacles, $\tilde{S}_\mu^0 =
\tilde{S}_{B_d(R)}(\mu;0)$ (given by Eq. (\ref{eq:tildeS_ball0})), as
a function of the dimensionless death rate $\mu R^2/D$ for small
obstacles ($r = 0.1$) {\bf (c)} and larger obstacles ($r = 0.2$) {\bf
(d)}.  Note that the difference is always positive, indicating that
reflecting obstacles inside a ball increase the exit time from the
ball, in agreement with the inequality (\ref{eq:Sineq}).}
\label{fig:discrepancy}
\end{figure}

Second, we consider the disk $B_2(R)$ with regularly distributed
reflecting circular obstacles, as illustrated in
Fig. \ref{fig:discrepancy}a,b.  Solving the problem numerically, we
obtain the Laplace-transformed survival probability in this
configuration and compare it with $\tilde{S}_{B_2}(\mu;0)$ from
Eq. (\ref{eq:tildeS_ball0}).  Figure \ref{fig:discrepancy}c,d shows
the relative difference between these two quantities, confirming again
that the obstacles contribute to an increase of the survival
probability.  In spite of various qualitative and numerical evidences,
our claim remains conjectural from a mathematical point of view.

\section{Escape from the disk through an arc on the boundary}
\label{sec:Adisk}

The exact solution for the Laplace-transformed survival probability
was obtained for immortal walkers in several planar domains such as a
disk, an angular sector, an annulus, and a rectangle
\cite{Rupprecht15}.  For the sake of completeness, we reproduce the
exact and approximate solutions for the Laplace-transformed survival
probability of a walker inside the disk of radius $R$ with the
reflecting boundary except for an escape arc $\Gamma = \{ (R,\theta)
~:~ \theta \in (\pi-\ve,\pi+\ve)\}$ of angle $2\ve$.  From these
expressions, we easily get the MFPT and the exit probability for
mortal walkers as discussed in the text.  We also investigate their
asymptotic behavior.

\subsection{Exact solution}

The Laplace-transformed survival probability reads \cite{Rupprecht15}
\begin{equation}
\label{eq:Sdisk_exact}
\tilde{S}(p;r_0,\theta_0) = \tilde{S}^\pi(p;r_0) + \frac{R^2}{D} ~ u\bigl(pR^2/D; r_0/R,\theta_0\bigr), 
\end{equation}
where
\begin{equation}
\tilde{S}^\pi(p;r_0) = \frac{1}{p}\biggl(1 - \frac{I_0(r_0 \sqrt{p/D})}{I_0(R\sqrt{p/D})}\biggr)
\end{equation}
is the rotation-invariant solution for the whole absorbing circle (see
Eq. (\ref{eq:uniform_mfpt_2D})), while the auxiliary function $u$
(which is positive because $S(t;x_0) \geq S_\pi(t;x_0)$) can be
expressed as
\begin{equation}
u(\p;~r,\theta) = \sum\limits_{n=0}^\infty a_n(\p) \frac{I_n(r \sqrt{\p})}{I_n(\sqrt{\p})} \cos(n\theta) ,
\end{equation}
where the coefficients $a_n(\p)$ are uniquely determined through the
boundary conditions (note that we replace the notation $a_0(\p)/2$ by
$a_0(\p)$ as compared to \cite{Rupprecht15}):
\begin{equation}
\begin{split}
a_0(\p) & = C_\ve(\p) \biggl[\frac{\sqrt{\p}~ I_0(\sqrt{\p})}{I_1(\sqrt{\p})} + \p\,C_\ve(\p)\biggr]^{-1},   \\
a_n(\p) & = 2\hat{a}_n(\p) \biggl[\frac{\sqrt{\p}~ I_0(\sqrt{\p})}{I_1(\sqrt{\p})} + \p\,C_\ve(\p)\biggr]^{-1},   \\
\end{split}
\end{equation}
where the auxiliary coefficients $\hat{a}_n(\p)$ are determined as
\begin{equation}
\label{eq:ahat}
\hat{a}_n(\p) = \biggl[(I - M \gamma)^{-1} \alpha \biggr]_n  \qquad (n \geq 1).
\end{equation}
In this relation, $I$ is the identity matrix, $\gamma$ is the diagonal
matrix formed by
\begin{equation}
\gamma_n(\p) = 1 - \frac{\sqrt{\p}}{2n}~ \frac{I_{n-1}(\sqrt{\p}) + I_{n+1}(\sqrt{\p})}{I_{n}(\sqrt{\p})}  \qquad (n \geq 1),
\end{equation}
$\alpha$ is the vector formed by 
\begin{equation}
\label{eq:alphan}
\alpha_n = \frac{(-1)^{n-1}}{2n} \biggl[P_{n-1}(\cos \ve) + P_n(\cos \ve)\biggr]  \qquad (n\geq 1)
\end{equation}
($P_n(z)$ being Legendre polynomials), the matrix $M$ is
\begin{equation}
M_{nm} = \frac{m}{2} \hspace*{-2mm} \int\limits_{-\cos \ve}^1 \hspace*{-2mm} \frac{dx}{1+x} \bigl[P_{m-1}(x) + P_m(x)\bigr] \bigl[P_{n-1}(x) + P_n(x)\bigr] 
\end{equation}
(for $m,n\geq 1$), and
\begin{equation}
\label{eq:Cp}
C_\ve(\p) = -2 \ln \sin(\ve/2) + 2\sum\limits_{n=1}^\infty n \alpha_n \hat{a}_n(\p) \gamma_n(\p) .
\end{equation}

From Eq. (\ref{eq:Sdisk_exact}), we deduce the exit probability
$H_\mu(x_0)$:
\begin{equation}
\label{eq:Hmu_disk}
\begin{split}
H_\mu(x_0) & = \biggl(\frac{\sqrt{s} \, I_0(\sqrt{s})}{I_1(\sqrt{s})} + s\, C_\ve(s)\biggr)^{-1} 
\biggl[\frac{\sqrt{s} \, I_0(\frac{r_0}{R}\sqrt{s})}{I_1(\sqrt{s})} \\
& -2s \sum\limits_{n=1}^\infty \hat{a}_n(s) \frac{I_n(\frac{r_0}{R}\sqrt{s})}{I_n(\sqrt{s})} \cos n\theta_0 \biggr] , \\
\end{split}
\end{equation}
with $s = \mu R^2/D$, $r_0 = |x_0|$, and $\theta_0$ being the angular
coordinate of the starting point $x_0$.  Setting $r_0 = 0$, one gets
\begin{equation}
\begin{split}
H_\mu(0) & = \biggl(\frac{\sqrt{s}\, I_0(\sqrt{s})}{I_1(\sqrt{s})} + s\, C_\ve(s)\biggr)^{-1} \frac{\sqrt{s}}{I_1(\sqrt{s})} , \\
\end{split}
\end{equation}
from which Eq. (\ref{eq:Hmu_disk}) can be re-written as
\begin{equation}
\label{eq:Hmu_disk2}
\begin{split}
H_\mu(x_0) & = H_\mu(0) \biggl[I_0\biggl(\frac{r_0}{R}\sqrt{s}\biggr) \\
& -2\sqrt{s} \, I_1(\sqrt{s}) \sum\limits_{n=1}^\infty \hat{a}_n(s) \frac{I_n(\frac{r_0}{R}\sqrt{s})}{I_n(\sqrt{s})} \cos n\theta_0 \biggr] . \\
\end{split}
\end{equation}
If the starting point $x_0$ is uniformly distributed over the disk,
the volume average yields the GEP
\begin{equation}
\label{eq:GEP_disk}
\begin{split}
\overline{H}_\mu & \equiv \frac{1}{\pi R^2} \int\limits_0^R dr_0 ~ r_0 \int\limits_0^{2\pi} d\theta_0 \, H_\mu(r_0,\theta_0) \\
& = \frac{2I_1(\sqrt{s})}{\sqrt{s}} \, H_\mu(0) = 2 \biggl(\frac{\sqrt{s}\, I_0(\sqrt{s})}{I_1(\sqrt{s})} + s\, C_\ve(s)\biggr)^{-1} . \\
\end{split}
\end{equation}
Finally, the volume-averaged probability density of the FPP from
Eq. (\ref{eq:omega_mu_MFPT}) reads 
\begin{equation}
\label{eq:omega_disk}
\overline{\omega}_\mu(\theta) = \frac{1}{\pi R} \biggl[\frac{I_1(\sqrt{s})}{\sqrt{s}\, I_0(\sqrt{s})} - \sqrt{s} 
\sum\limits_{n=0}^\infty \hat{a}_n(s) \frac{I'_n(\sqrt{s})}{I_n(\sqrt{s})} \cos n\theta_0 \biggr] .
\end{equation}

Note that if the escape region is the whole boundary (i.e., $\ve =
\pi$), all $\alpha_n = 0$ that implies $C_\pi(s) = 0$ and $a_n(s) =
\hat{a}_n(s) = 0$ for $n \geq 1$.  In particular, one gets 
\begin{equation}
\label{eq:GEP_2dwhole}
\overline{H}_\mu \simeq \frac{2}{R \sqrt{\mu/D}} \qquad (\mu \gg D/R^2) ,
\end{equation}
in agreement with the general asymptotic relation
(\ref{eq:Hmu_limit}).

Moreover, if now the relation (\ref{eq:Hmu_limit}) is applied for a
given size $\ve$ of the escape region,
\begin{equation}
\overline{H}_\mu \simeq \frac{2\ve/\pi }{R \sqrt{\mu/D}} \qquad (\mu \gg D/R^2) ,
\end{equation}
Eq. (\ref{eq:GEP_disk}) implies the asymptotic behavior for $H_\mu(0)$
\begin{equation}
\label{eq:Hmu0_disk}
H_\mu(0) \simeq \frac{\ve \sqrt{2}}{\sqrt{\pi}} \, (R^2 \mu/D)^{\frac 14} \, e^{-R\sqrt{\mu/D}}  \qquad (\mu \gg D/R^2) .
\end{equation}
Interestingly, this behavior is different from the case of the
concentric escape region, for which Eq. (\ref{eq:exit_2D}) yields for
$r=R$:
\begin{align} 
H_\mu(R) \underset{\mu \rightarrow \infty}{\sim} \frac{\sqrt{2\pi}\,  e^{-R \sqrt{\mu/D}} }
{(R^2\mu/D)^{\frac 14} \biggl(\gamma + \ln \frac{\ve  R\sqrt{\mu/D}}{2\pi} \biggr)} ,
\end{align}
where the radius $a$ of the escape region was expressed as $a = R \ve
/\pi$ so get the same perimeter of the escape region.  The dependence
of the exit probability on the size of the escape region is very
different in both cases.

\subsection{Approximate solution for small $\ve$}

The above solution given by Eqs. (\ref{eq:Sdisk_exact} --
\ref{eq:GEP_disk}) is exact and valid for any size $2\ve$ of the
escape region.  However, this solution is not explicit as the
inversion of the infinite-dimensional matrix in Eq. (\ref{eq:ahat}) is
needed.  In practice, the matrices $M$ and $\gamma$ can be truncated
to a finite size and then inverted numerically, leading to the
solution with any desired accuracy.  

For small $\ve$, the matrix $M$ was shown to be close to the identity,
yielding an explicit approximate solution,
\begin{eqnarray}
\label{eq:hata_approx}
\hat{a}_n(\p) & \simeq &\frac{\alpha_n}{1 - \gamma_n(\p)} , \\
\label{eq:Capprox}
C(\p) & \simeq & -2 \ln \sin(\ve/2) + 2\sum\limits_{n=1}^\infty \frac{n \alpha_n^2 \gamma_n(\p)}{1-\gamma_n(\p)} ,
\end{eqnarray}
that was shown to be very accurate for $\ve$ small enough
\cite{Rupprecht15}.  In particular, the approximate expression for
$C(\p)$ determines {\it explicitly} the global exit probability
according to Eq. (\ref{eq:GEP_disk}).

Using these relations, one can investigate the asymptotic behavior of
the survival probability, the MFPT, and the exit probability
$H_\mu(x_0)$ in the narrow escape limit ($\ve \ll 1$).  Here, we focus
only on the global exit probability $\overline{H}_\mu$ as the most
important safety characteristic of a container.  For small $\mu$, one
has $\gamma_n(\p) \simeq - \p/(2n(n+1))$, from which
\begin{equation}
C(\p) \simeq 2\ln(2/\ve) + O(\p),
\end{equation}
and Eq. (\ref{eq:GEP_disk}) implies
\begin{equation}
\overline{H}_\mu \simeq \frac{1}{1 + (\mu R^2/D) \ln (2/\ve)} .
\end{equation}
Given that $\langle\tau_0\rangle \simeq \frac{R^2}{D} \ln(2/\ve)$,
this expression agrees with our general approximation
(\ref{eq:Hmu_approx}) for intermediate death rates.

\subsection{Quality of the upper bound (\ref{eq:H_upper})}
\label{sec:AHupper}

\begin{figure}
\begin{center}
\includegraphics[width=80mm]{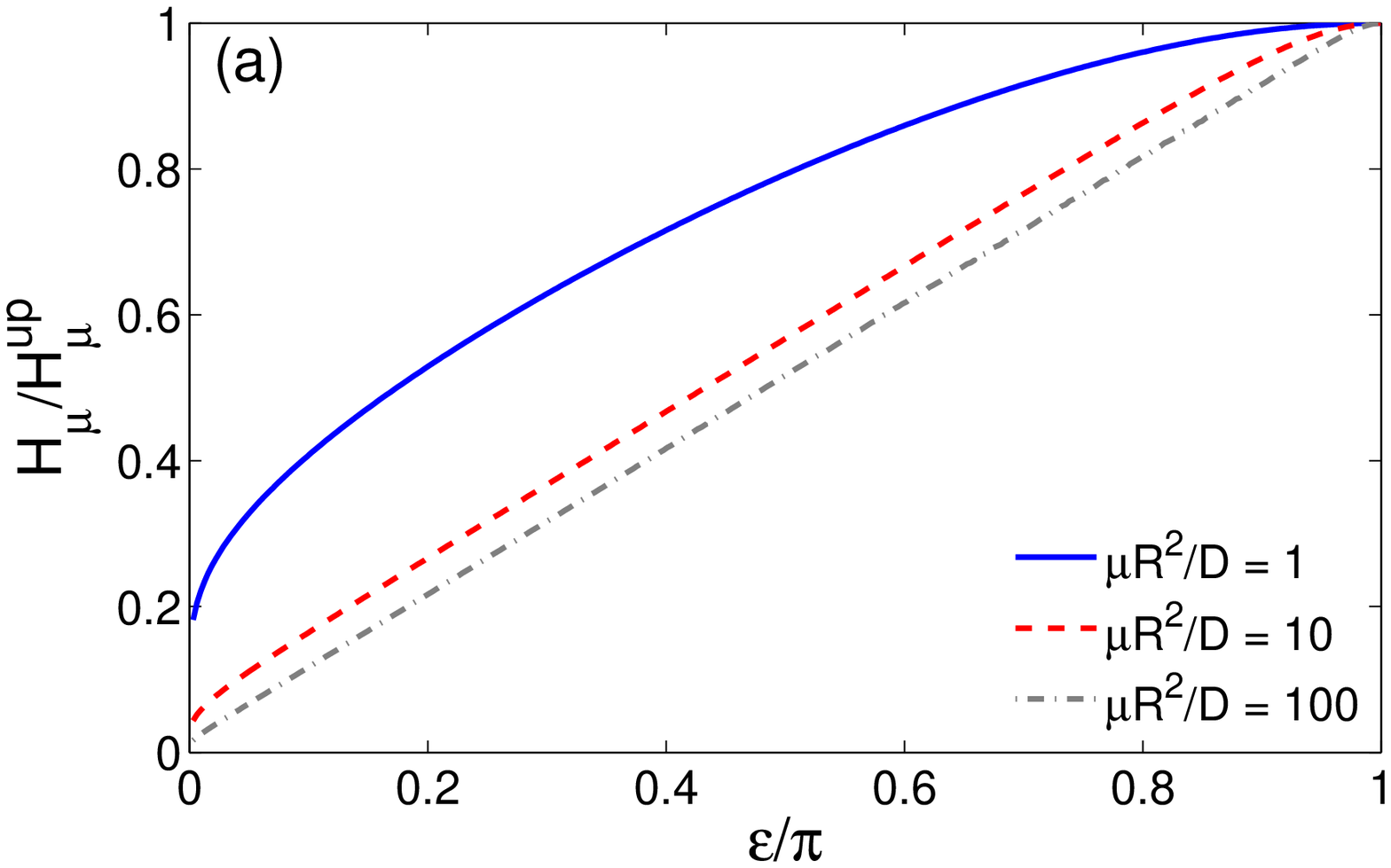} 
\includegraphics[width=80mm]{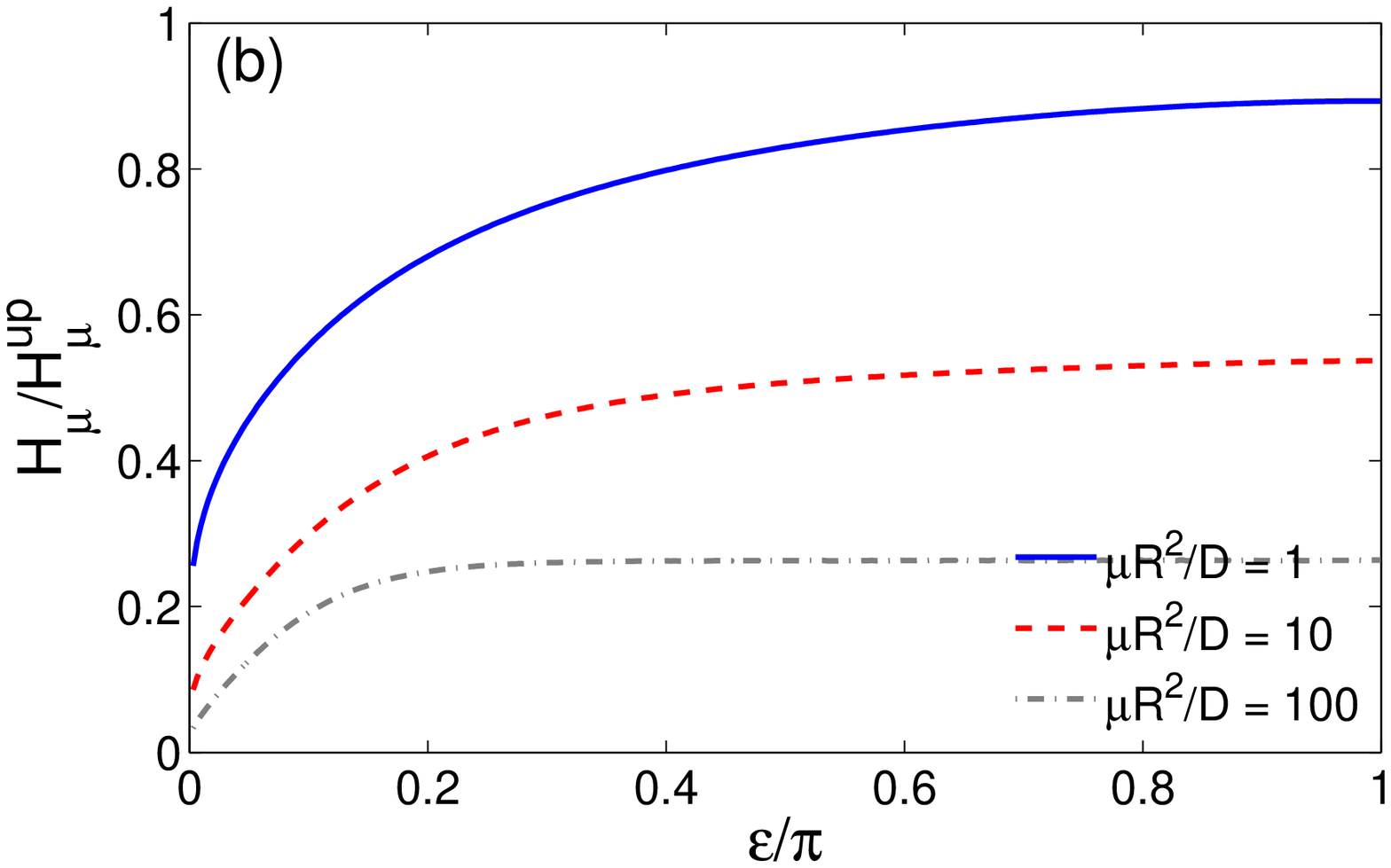} 
\end{center}
\caption{
(Color online) The exit probability $H_\mu(x_0)$ (given by
Eq. (\ref{eq:Hmu_disk})), normalized by its upper bound $H_\mu^{\rm
up}(x_0) = \U_2(|x_0-\pa|\sqrt{\mu/D})$ from
Eq. (\ref{eq:H_upperdef}), as a function of the normalized perimeter
$\ve/\pi$ of the escape arc $(\pi-\ve,\pi+\ve)$, for the disk of
radius $R$ at several values of $\mu$.  We set $x_0 = 0$ {\bf (a)} and
$x_0 = -0.5$ (i.e., $r_0 = 0.5$ and $\theta_0 = \pi$) {\bf (b)}. }
\label{fig:Hmu_0_epsall}
\end{figure}

In this section, we examine the quality of the upper bound
(\ref{eq:H_upper}) in the case of a disk.  Figure
\ref{fig:Hmu_0_epsall}a shows the exit probability $H_\mu(0)$ for
the disk, normalized by the upper bound $\U_2(R\sqrt{\mu/D})$, as a
function of $\ve$, for several values of the death rate $\mu$.  One
can see that, at large $\mu$, this ratio becomes proportional to
$\ve/\pi$.  This result agrees with the following qualitative picture:
at large $\mu$, the arrival onto the boundary is a rare event; only
short trajectories that rapidly reach the boundary, can provide a
notable contribution.  If such a trajectory arrives onto the
reflecting part of the boundary, it needs extra time to diffuse
towards the escape region, and thus provides a much smaller
contribution.  As a consequence, the overall probability to reach the
boundary, which is given by the upper bound, is multiplied by the
fraction of trajectories arriving to the escape region, i.e.,
$\ve/\pi$.

When the boundary points are not equally distant from the starting
point, their contributions are not equal, and more complicated
dependence on the escape region size is expected.  This is illustrated
in Fig. \ref{fig:Hmu_0_epsall}b showing the ratio
$H_\mu(x_0)/H_\mu^{\rm up}(x_0)$ at the starting point $x_0 = -0.5$
(i.e., $r_0 = 0.5$ and $\theta_0 = \pi$) which is closer to the escape
region.  This ratio first grows proportionally to $\ve$ (at small
$\ve$) but then saturates to a constant because any further addition
of distant boundary points does not facilitate the escape from the
disk at large $\mu$.

\section{Upper bound of the survival probability for the whole absorbing boundary}
\label{sec:A_upper}

The proposed upper bound (\ref{eq:Supper}) of the Laplace-transformed
survival probability, which follows from a trivial inequality
$S(t,x_0)\leq 1$, implies a useless lower bound for the exit
probability: $H_\mu(x_0) \geq 0$.  Finding more informative lower
bounds for $H_\mu(x_0)$ is an interesting perspective.  In this
Appendix, we propose one improvement which, however, is only valid in
the case when the escape region is the whole boundary: $\Gamma =
\pa$.

The domain monotonicity for Dirichlet heat kernels \cite{Davies}
implies the general upper bound for the survival probability in a
domain $\Omega \subset \R^d$ with fully reactive boundary $\pa$:
\begin{equation}
\label{eq:Stupper}
S(t;x_0) \leq \int\limits_\Omega dx ~ \frac{\exp\bigl(-\frac{|x-x_0|^2}{4Dt}\bigr)}{(4\pi Dt)^{d/2}} ,
\end{equation}
where the Gaussian propagator is explicitly written on the right-hand
side.  Splitting the integral into two contributions, from the ball of
radius $|x_0-\pa|$ centered at $x_0$, and the rest, one gets
\begin{equation}
\begin{split}
S(t;x_0) & \leq 1 - \frac{\Gamma\bigl(\frac{d}{2};~ |x_0-\pa|^2/(4Dt)\bigr)}{\Gamma(\frac{d}{2})} \\
& +  \biggl(|\Omega| - \frac{\pi^{\frac{d}{2}} |x_0-\pa|^d}{\Gamma(\frac{d}{2}+1)} \biggr)
\frac{\exp\bigl(-\frac{|x_0-\pa|^2}{4Dt}\bigr)}{(4\pi Dt)^{\frac{d}{2}}} , \\
\end{split}
\end{equation}
where $\Gamma(\nu,z)$ is the upper incomplete Gamma function, and
$|\Omega|$ is the volume of $\Omega$ from which the volume of the ball
of radius $|x_0-\pa|$ is subtracted.  The Laplace transform reads
\begin{equation}
\label{eq:S_upper1}
\begin{split}
\tilde{S}(p;x_0) & \leq \frac{1}{p} \biggl[1 - \frac{2^{1-\frac{d}{2}}}{\Gamma(\frac{d}{2})} 
s^{\frac{d}{4}} K_{\frac{d}{2}}\bigl(\sqrt{s}\bigr) \\
& + 2^{-\frac{d}{2}}  \biggl(\frac{|\Omega|}{\pi^{\frac{d}{2}} L^d} - \frac{1}{\Gamma(\frac{d}{2}+1)} \biggr) 
s^{\frac{d+2}{4}} K_{1-\frac{d}{2}}\bigl(\sqrt{s}\bigr) \biggr], \\
\end{split}
\end{equation}
where $s = |x_0-\pa|^2 p/D$, and we used formula 6.453 from
\cite{Gradshteyn} for the first Laplace transform involving the
incomplete Gamma function.  Since $H_\mu(x_0) = 1-\mu
\tilde{S}(\mu;x_0)$, this inequality implies a lower bound for the
exit probability $H_\mu(x_0)$.  At large $s$, the right-hand side
becomes
\begin{equation}
\begin{split}
\tilde{S}(p;x_0) \lesssim \frac{1}{p} \biggl[1 & - \frac{\sqrt{\pi}\, 2^{-\frac{d+1}{2}}}{ \Gamma(\frac{d}{2}+1)} \, s^{\frac{d-1}{4}} \, e^{-\sqrt{s}} \\
& \times \biggl(d - \biggl(\frac{|\Omega| \Gamma(\frac{d}{2}+1)}{\pi^{\frac{d}{2}} L^d} - 1 \biggr) \sqrt{s} \biggr) \biggr]. \\
\end{split}
\end{equation}
One can see that the correction to the leading $1/p$ term vanishes as
$e^{-\sqrt{s}}$.  However, at large $s$, the correction term becomes
positive and thus useless because of the trivial upper bound
$\tilde{S}(p;x_0) \leq 1/p$.  In other words, the upper bound
(\ref{eq:S_upper1}) improves the trivial upper bound $1/p$ only for
moderate values of $p$ (or $s$).

We recall that the upper bound (\ref{eq:S_upper1}) is only applicable
for the whole absorbing boundary ($\Gamma = \pa$), while its extension
to mixed Dirichlet-Neumann boundary condition (with $\Gamma \ne \pa$)
is not valid in general.  In fact, the domain monotonicity does not
hold in general for Neumann heat kernels \cite{Davies}.  For instance,
in the limit of no escape region, the survival probability is equal to
$1$, and the inequality (\ref{eq:Stupper}) and its consequence
(\ref{eq:S_upper1}) do not hold.  Finding more accurate upper and
lower bounds for the Laplace-transformed survival probability for the
escape problem presents an important direction for future research.

\section{Marginal probability densities for the FPT and FPP}
\label{sec:A_omega}

In this Appendix, we deduce Eqs. (\ref{eq:rho_omega},
\ref{eq:omega_mu}).  The first equation is simply obtained as 
\begin{eqnarray} 
\nonumber
\rho(t;x_0) &=&  - \frac{\partial}{\partial t} S(t;x_0) = - \frac{\partial}{\partial t} \int\limits_\Omega dx \,  G_t(x,x_0) \\
\nonumber
&=& -\int\limits_\Omega dx \, D \Delta G_t(x,x_0) = - D
\int\limits_{\pa} dx \, \frac{\partial}{\partial n} G_t(x,x_0) \\ 
&=& \int\limits_\Gamma dx \, q(t,x;x_0) ,
\end{eqnarray}
where we used Eqs. (\ref{eq:S_G}, \ref{eq:rho}), the divergence
theorem, the diffusion equation, and boundary conditions.  

To deduce Eq. (\ref{eq:omega_mu}), we note that, in analogy to the
harmonic measure density, the marginal probability density
$\omega_\mu(y;x)$ for the FPP, $y\in \Gamma$, satisfies the following
PDE:
\begin{equation}
\label{eq:Aomega_mu}
\begin{split}
\bigl[D \Delta -\mu\bigr] \omega_\mu(y;x) & = 0 \quad (x\in\Omega), \\
\omega_\mu(y;x) & = \delta(x-y) \quad (x\in \Gamma) , \\
\frac{\partial}{\partial n} \omega_\mu(y;x) & = 0 \quad (x \in \pa\backslash \Gamma) . \\
\end{split}
\end{equation}
The density $\omega_\mu(y;x)$ can be expressed through the
Laplace-transformed propagator $\tilde{G}_\mu(x;x_0)$ which satisfies
\begin{equation}
\label{eq:AG_mu}
\begin{split}
\bigl[D \Delta -\mu\bigr] \tilde{G}_\mu(x;x_0) & = -\delta(x-x_0) \quad (x\in\Omega), \\
\tilde{G}_\mu(x;x_0) & = 0 \quad (x\in \Gamma) , \\
\frac{\partial}{\partial n} \tilde{G}_\mu(x;x_0) & = 0 \quad (x \in \pa\backslash \Gamma) . \\
\end{split}
\end{equation}
Multiplying the first equation in (\ref{eq:Aomega_mu}) by
$\tilde{G}_\mu(x;x_0)$, the first equation in (\ref{eq:AG_mu}) by
$\omega_\mu(y;x)$, subtracting them and integrating over $x\in\Omega$,
one gets
\begin{equation}
\begin{split}
& \omega_\mu(y;x_0) \\
& = D \int\limits_\Omega dx \biggl[\tilde{G}_\mu(x;x_0) \Delta \omega_\mu(y;x) - \omega_\mu(y;x) \Delta \tilde{G}_\mu(x;x_0) \biggr] \\
& = -D \int\limits_{\pa} dx \biggl[\tilde{G}_\mu(x;x_0) \frac{\partial \omega_\mu(y;x)}{\partial n} 
- \omega_\mu(y;x) \frac{\partial \tilde{G}_\mu(x;x_0)}{\partial n}  \biggr] \\
& = - D \frac{\partial}{\partial n} \tilde{G}_\mu(x;x_0) = \int\limits_0^\infty dt \, e^{-\mu t} \, q(t,x;x_0), \\
\end{split}
\end{equation}
where we used the Green formula and the boundary conditions for both
$\tilde{G}_\mu(x;x_0)$ and $\omega_\mu(y;x)$.  When $\mu = 0$, one
retrieves the harmonic measure density, i.e., the probability density
of the FPP for immortal walkers.


\end{document}